\documentclass[aps,twocolumn,showpacs,showkeys,amsmath,amssymb,superscriptaddress,floatfix]{revtex4}
\usepackage{graphicx}
\usepackage{dcolumn}
\usepackage{bm}
\begin{document}
\title{Linear sigma model at finite temperature}
\author{Nicholas Petropoulos}
\affiliation{Department of Physics and
Astronomy, University of Manchester, Manchester M13 9PL, United Kingdom} 

\affiliation {Centro de F\'{\i}sica Te\'{o}rica, Departamento de
F\'{\i}sica, Universidade de Coimbra, P3004-516 Coimbra, Portugal}

\date{February 2004\footnote{This work is based on a thesis report with title ``Linear sigma model and chiral symmetry
at finite temperature'', which
was submitted as part of the requirements
for a PhD degree in Manchester University in
September 2000.  Some
new references and comments have been added. Some of the results presented in 
this work  have  been 
published already in \cite{Petropoulos:1999gt,Petropoulos:1998en}. 
}}

\email{nicholas@teor.fis.uc.pt}

\begin{abstract}
The chiral phase transition
is  investigated within the framework of 
thermal field theory using the linear sigma model as an effective theory. We  concentrate on
the meson sector of the model, and calculate the thermal 
effective potential in the Hartree approximation 
by using the Cornwall--Jackiw--Tomboulis formalism of composite
operators. The thermal effective potential is calculated for $N=4$ involving 
as usual the sigma and the three pions, and in the large--$N$ 
approximation involving $N-1$ pion fields. In the $N=4$ case, we have
examined the theory both in the chiral limit and with the presence
of a symmetry breaking term, which is responsible for the generation of the pion
masses. In both cases, the system 
of the resulting  gap equations
for the thermal effective masses of the particles has been solved numerically, and we have 
investigated the evolution of the effective 
potential. In the $N=4$ case, there is indication
of a first--order phase transition, and the Goldstone
theorem is not satisfied. The situation is
different in the general case where we have  used the large--$N$ approximation. The 
Goldstone theorem is satisfied, and the phase transition  
appears as second--order. In our analysis, we have ignored quantum
fluctuations and have used the imaginary time formalism for calculations.
We extended our calculation in order to include the full effect of two
loops in the calculation of the effective potential. In this case, the 
effective masses are momentum dependent. In order to perform
the calculations, we found the real time formalism to be 
convenient. We have calculated the effective 
masses of pions at the low--temperature phase and we found a quadratic
dependence on temperature, in contrast to the Hartree case, where
the mass is proportional to temperature. The sigma mass 
was investigated in the presence 
of massive pions, and we found a small deviation compared to the Hartree
case. In all cases, the system approaches the behaviour of the 
ideal gas at the high temperature limit.
\end{abstract}

\maketitle

\tableofcontents

\section{\label{Sect:First}Chiral symmetry breaking and restoration }

\subsection{Overview} 

The study of matter at very high  temperatures and densities 
and of the phase transitions which take place between  
the different phases, has several very interesting  
aspects. It has been the subject 
of intense study during the last few years, because of its relevance 
to  particle  
physics, astrophysics and cosmology. According to the standard 
big bang model, it is believed that a series of phase  
transitions happened at  
the early stages of the evolution of the universe, the QCD phase 
transition being one of them
\cite{Mclerran:1986zb,Gross:1981br,Smilga:1997cm,Linde:1979px,Linde:1990,Rajagopal:1995bc}. 

There is hope at present that it may also be  
possible to probe this transition in the laboratory 
in experiments involving 
relativistic heavy ion collisions. Experiments
of this type carried out at CERN may have 
reached the phase transition \cite{Heinz:2000ba}. Further experiments 
are planned in the future, and the results
could possibly improve our knowledge
on subjects such as the restoration of chiral symmetry, the nature 
of the quark gluon plasma and the color superconducting phase, as well as the 
physics of neutron and quark stars \cite{Rischke:2003mt,Ruster:2003zh}.

There are two main issues related to the QCD phase 
transition, namely restoration of chiral symmetry (chiral 
phase transition), and deconfinement of quarks and gluons
to form  the so called quark--gluon plasma. At present, it is 
not at all clear what the relation is between these two
phase transitions, if they happen at the same temperature, or 
if they are independent. Another issue is the order of this
transition, is it first--order with latent heat, or 
second--order, or maybe a crossover between the phases? Lattice 
calculations suggest that when we consider two massless quarks, the 
transition is second order and the same suggest
other approaches based on effective models. If we consider
three massless flavours of quarks,  the transition 
is probably first--order
\cite{Smilga:1997cm,Rajagopal:1995bc}. The aim of this work is to study the 
chiral phase transition.

Chiral symmetry breaking is a necessary ingredient
for low energy hadron physics, since unbroken chiral symmetry 
results in  massless baryons, without 
parity doubled partners. It is well known that there is no
parity partner of the proton  and that the proton is not 
massless, therefore chiral symmetry must be broken. However,
any case in which a global symmetry is broken 
gives rise to the appearance of massless  
Goldstone bosons. In reality, there are no  massless particles in the
hadron spectrum but the pions are very light, so one could consider them as approximate 
Goldstone bosons. It is generally believed that at some temperature, or baryon density, the 
chiral symmetry could be restored. 

An important part when studying questions like  
the restoration of spontaneously broken  
symmetries is the construction of order parameters, which 
characterise the way in which the symmetry of the system 
under consideration is realised. These quantities are zero  
in the phase where the symmetry is manifest but non zero in the 
spontaneously broken phase. A classic example of an order parameter is the 
magnetisation of a ferromagnetic substance, which 
is non--zero below  the Curie temperature, but  
disappears at  higher temperatures. The system 
undergoes a transition from an asymmetric  ordered  
state, with non--zero magnetisation at low temperature  to  a  
symmetric disordered  state, with zero magnetisation 
at temperatures 
well above the Curie point. We usually encounter two types of  
phase transition. In first--order transitions, the 
order parameter jumps discontinuously from its value in  
one phase to that in the other (usually zero). In contrast, during  
second order transitions, the order parameter  
vanishes continuously \cite{Linde:1979px,Linde:1990}.

An important order parameter for the chiral phase 
transition of QCD is the  
quark condensate, a measure of the density of quark-antiquark pairs  
that have condensed into the same 
quantum mechanical state. They fill the lowest energy state 
-- the vacuum of QCD --  
and as a result the chiral symmetry  
is broken, since there is no invariance under 
chiral transformations \cite{Birse:1994cz,Hatsuda:2000by,Kunihiro:2000je}.
It is expected  that at very high temperatures, the quark 
condensate disappears, and the system 
is  chirally symmetric. 

There are two main paths to study the chiral phase transition, namely
lattice QCD methods and effective field theories. Lattice QCD can tell
us many things about the phase transition, but cannot be used to study
its dynamics. Hence, the need for models which do not suffer
this restriction. In the case of chiral symmetry, a model with 
the correct chiral properties is  
the linear sigma model, a theory of 
fermions (quarks or nucleons), interacting with 
mesons \cite{Gell-Mann:1960np}. This model has been used extensively  
as an effective theory 
in the low energy phenomenology of QCD, describing the physics of  
mesons, and it is well suited for 
a study of the chiral phase 
transition \cite{Rajagopal:1995bc,Birse:1994cz}. In particular the
model is very popular in studies of the so--called 
disoriented chiral condensates (DCC). We review the model and discuss 
how it is related   
to chiral symmetry in Section~\ref{Sect:Chiral}. For this presentation we have 
followed closely the lecture notes on spontaneous 
breaking of chiral symmetry by Li \cite{Li:1999yx}. 
 
The appropriate 
framework for study of phase transitions is thermal field theory 
or finite temperature 
field theory, a combination of quantum 
field theory and statistical mechanics. 
Within this framework, the finite--temperature 
effective potential is an 
important and well--used theoretical tool. The use of such 
techniques 
goes back to the 1970's when  
Kirzhnits and Linde \cite{Kirzhnits:1972iw,Kirzhnits:1972ut}  
first proposed that  symmetries broken at zero temperature  
could be restored at finite 
temperatures. Subsequent work  
by Weinberg \cite{Weinberg:1974hy}, Dolan and Jackiw \cite{Dolan:1974qd}, 
as well as many others, resulted in a wide adoption  
of the effective potential  
as the basic tool in such studies. The 
basic ideas about the notion of the effective potential are reviewed 
in the Section~\ref{Sect:Potential}. However, a detailed analysis can be found 
in the original papers, and also in recent textbooks on the subject
\cite{Rivers:1987,Kapusta:1989,LeBellac:1996,Das:1997gg}.
  
The finite--temperature effective potential  
$V(\phi,T)$  is defined through an effective  
action $\Gamma (\phi)$,  which is the  
generating functional  of the one--particle irreducible  
graphs, and it is related to  the free energy of 
the system \cite{Dolan:1974qd}. A generalised version is the   
effective potential $V(\phi,G)$ for 
composite operators introduced by Cornwall, Jackiw and  
Tomboulis (CJT) \cite{Cornwall:1974vz} and later
formulated at finite temperature by Amelino--Camelia and Pi
\cite{Amelino-Camelia:1993nc,Amelino-Camelia:1994kd,Amelino-Camelia:1996hw}. 
The
composite operator method in calculating thermodynamic
potentials for many--body systems, was also introduced
in the 1960s by Luttinger and Ward \cite{Luttinger:1960} 
(see also \cite{Baym:1962}), and has recently been  used
in studies of systems in and out of equilibrium
\cite{vanHees:2000bp,Ivanov:2000ma}.

When one 
is working in the context of field theory at finite 
temperature, there are basically three main paths to follow, 
namely the imaginary time formalism, the real time 
formalism and thermo--field dynamics \cite{Das:1997gg}. Which of these is 
the more convenient depends on the application at hand. 
There are advantages and disadvantages to each of the three 
methods. We do not use thermo--field dynamical methods in 
what follows, but we will use both real and imaginary time 
formalisms. We give a few basics about the two types of formalism in  
Section~\ref{Sect:Formalisms}. 
Finally in  Section~\ref{Sect:F4} we present  the  
calculation of the effective 
potential for the $\lambda \phi^{4}$ theory, using the CJT 
method.  

The outline of the remaining sections is as follows: 
in Section~\ref{Sect:Hartree}  we apply the CJT method  
to calculate the effective potential for the 
$O(4)$ linear sigma model 
in the Hartree approximation. We numerically solve the resultant system  
of gap equations, both in the chiral limit and with the presence of 
a linear term, which breaks the chiral symmetry  
of the Lagrangian. We have repeated these steps for the  
generalised version of the linear sigma model, with $N-1$ pion fields   
in the large--$N$ approximation, in Section~\ref{Sect:LargeN} . Then in   
Section~\ref{Sect:Beyond}, we extend our Hartree approximation for the $O(4)$ 
model in order to include the effects of the so--called 
sunset diagrams. In these calculations, we find the 
real time formalism more convenient.

In Section~\ref{Sect:Others}, we review recent attempts to incorporate the linear
sigma model and/or the CJT formalism in order to investigate
the physics of the hot pion gas, and  the chiral phase transition.
Finally, in Section~\ref{Sect:Summary}, we outline our conclusions.

\subsection{\label{Sect:Chiral}Chiral symmetry}

The symmetry principle is possibly one of the most important ideas in the
development of high energy physics. As it is well known, the symmetries of a
physical system lead to conservation laws, which have as consequence  many important
relations for the physical processes. However, many of the symmetries which we observe in nature are
approximate symmetries rather than exact ones. A very interesting mechanism 
which is related to these ideas 
is the spontaneous symmetry breaking (SSB) and it seems that  it has played an
individual role in the development and our present understanding 
of particle physics. The main characteristic of spontaneous symmetry breaking 
is the fact that it is related to the ground state of the theory. The idea of SSB 
was first appeared  around 1960 in  studies
of superconductivity in  solid state physics by Nambu and Goldstone 
\cite{Nambu:1961tp,Goldstone:1961eq}. One of the important 
consequences of SSB is 
the presence of 
massless excitations which are reffered to as  the Nambu--Goldstone 
bosons.  Later, these  ideas were implemented in
particle physics. In combination with $SU(3)\times SU(3)$ current
algebra, SSB has been quite successful in our  understanding of the chiral
symmetry, a important ingredient of  the low energy phenomenology of the strong 
interactions. However, although SSB has been quite
successful in explaining many interesting phenomena, the implementation of this symmetry
the theoretical framework has been done rather arbitrarily, and  the 
origin of SSB is not absolutely clear.

A popular model which implements the ideas of chiral symmetry is
the linear sigma model which has a long and interesting 
history. It was originally
constructed in the 1960's, as a model to study the chiral symmetry in the
pion--nucleon  system \cite{Gell-Mann:1960np}. Later the spontaneous symmetry
breaking and PCAC (partially conserved axial current) were 
incorporated. Although this model is not quite phenomenologically 
correct, it remains an attractive 
example with interesting features, which displays many important aspects 
of broken symmetries. Nowdays it is a common belief that  the strong interactions 
are the best described by QCD, however the linear  sigma model, and also various
quark models, still serve  as effective theories at 
low energies, since at this regime due to confinenement  
it is difficult to calculate directly from QCD.

The Lagrangian for the linear sigma 
model is given by 
\begin{eqnarray}
\mathcal{L} &=&\frac{1}{2}\left( \partial _{\mu}\sigma \right)^{2} 
+\frac{1}{2}\left( \partial _{\mu} \bm{\pi}\right)^{2}
-\frac{\mu^{2}}{2}\left( \sigma^{2}+ \bm{\pi}^{2}\right) \\ \nonumber
&&-\frac{\lambda }{24}\left( \sigma ^{2}+\bm{\pi}^{2}\right)^{2}
+\bar{\psi}i\gamma ^{\mu}\partial _{\mu}\psi \\ \nonumber
&&+ig\bar{\psi}\bm{\tau}\gamma_{5}\psi \cdot \bm{\pi} 
+g\bar{\psi}\psi \sigma~,
\label{Eq:Full-Lagrangian}  
\end{eqnarray}
\noindent
where  the spinor field $\psi$ is  a massless isodoublet 
nucleon (or quark)  field. The pseudoscalar field 
$\bm{\pi }=( \pi _{1},\pi _{2},\pi_{3})$ is an isotriplet of pion fields
and $\sigma $ is the isosinglet
field.

This Lagrangian is now clearly invariant under the 
(infinitesimal) $SU(2)$ transformations, 
\begin{equation}
\delta\psi=-i\bm{\varepsilon}\cdot \bm{\tau}/2\psi,\qquad
\delta\bm{\pi}=\bm{\varepsilon}\times \bm{\pi},\qquad
\delta\sigma=0~,
\end{equation}
and the associated current is
\begin{equation}
J^{a \mu }(x)=(\pi \times\partial ^{\mu}\pi) ^{a}+\frac{1}{2}
\bar{\psi}\gamma ^{\mu }\tau ^{a}\psi~.
\end{equation}
We can verify
that the above Lagrangian is also invariant
under a further set of transformations, namely
\begin{equation}
\delta\psi=-i\bm{\eta}\cdot \bm{\tau}/2\gamma_{5}\psi,\qquad
\delta\pi=\bm{\eta} \sigma,\qquad
\delta\sigma=-\bm{\eta}\cdot\bm{\pi}~.
\label{Eq:Transformation}
\end{equation}
Then we have 
another set of conserved currents
\begin{equation}
J_{5}^{a\mu}=(\sigma \partial^{\mu }\pi^{a}
- \pi^{a}\partial^{\mu}\sigma ) +\frac{1}{2}
\bar{\psi}\gamma ^{\mu}\gamma_{5}\tau^{a}\psi~,
\label{Eq:Axial-Current}
\end{equation}
and the corresponding charges are given by
\begin{equation}
Q_{a}=\int d^{3}x J^{a 0}(x),\qquad Q_{5}^{a}=\int d^{3}x J_{5}^{a 0}~.
\end{equation}
A point that should be made here, is that 
the transformations given in Eq.~(\ref{Eq:Transformation}), involve changes in parity
and as a consequence the current in Eq.~(\ref{Eq:Axial-Current}) is an 
axial vector and the associated charges are pseudoscalars. Using the 
canonical commutation relations, we can derive the algebra
generated by those charges
\begin{subequations}
\begin{eqnarray}
\big [ Q^{a},Q^{b}\big ] &=& i\varepsilon _{abc}Q^{c}~,\\ 
\big [ Q^{a},Q_{5}^{b}\big ] &=& i\varepsilon _{abc}Q_{5}^{c}~,\\ 
\big [ Q_{5}^{a},Q_{5}^{b}\big ] &=& 0~,
\end{eqnarray}
\end{subequations}
which show that the $Q^a$ are the generators of an $SU(2)$, and that
the $Q^a_5$'s transform as an isovector under this $SU(2)$. We can 
now consider the combinations $Q^a_L$ and $Q^a_R$ defined as
\begin{equation}
Q_{L}^{a}=\frac{1}{2}(Q^{a}+Q^{a}_{5}), \qquad 
Q_{R}^{a}=\frac{1}{2}(Q^{a}-Q^{a}_{5})
\end{equation}
\noindent
to verify that they satisfy the following relations,
\begin{eqnarray}
&& \big [ Q_{L}^{a},Q_{L}^{b}\big ] =i\varepsilon _{abc}Q_{L}^{c}~,\\ \nonumber
&& \big [ Q_{R}^{a},Q_{R}^{b}\big ] =i\varepsilon _{abc}Q_{R}^{c}~,\\ \nonumber
&& \big [ Q_{R}^{a},Q_{L}^{b}\big ] =0~.
\end{eqnarray}
\noindent
Therefore $Q^a_L$ and $Q^a_R$ commute with each other, and they generate 
separate $SU(2)$ algebras. Then this combined algebra
is called $SU_{L}(2) \times SU_{R}(2)$.

There is also another way to describe the symmetry of the linear sigma
model, that  is the $O(4)$ symmetry which has the property of being isomorphic to 
$SU(2)\times SU(2)$. If we apply the transformation
\begin{equation}
R_{ij}=\delta _{ij}+\varepsilon _{ij},\qquad \textrm{with}\qquad 
\varepsilon_{ij}=-\varepsilon _{ji}~, 
\end{equation}
with $R$  an orthogonal  $4\times 4$ matrix, then the scalar field 
$\phi_i=(\sigma, \pi_1,\pi_2,\pi_3)$ transforms as 
4--dimensional vector, 
\begin{equation}
\phi _{i}\longrightarrow \phi _i '=R_{ij}\phi _j\simeq \phi_i+\varepsilon _{ij}\phi _j~. 
\end{equation}
We then observe that the combination $\phi_i^2 = \sigma^2+\bm{\pi}^2$, is just the length 
of the vector $\phi _{i}$, and is clearly invariant
under the rotations in 4--dimensions. If we take 
\begin{equation}
\varepsilon _{ij}=\varepsilon _{ijk}\alpha _k,
\qquad \varepsilon_{4i}=\beta _i,\qquad i,j,k=1,2,3 
\end{equation}
\noindent
we find that 
\begin{equation}
\bm{\pi '}=\bm{\pi}+
\bm{\alpha}\times \bm{\pi}+\bm{\beta}\sigma ,
\qquad \sigma '=\sigma+ \bm{\beta}\cdot \bm{\pi}~. 
\end{equation}
Therefore we see that the parameters 
$\bm{\alpha}$, correspond to a vector transformation, while
$\bm{\beta}$ to an axial transformation. This is a very important 
property, since this $O(4)$ symmetry can easily 
be generalised to $O(N)$ one, involving  $N$ pion fields.  

\subsubsection{Spontaneous symmetry breaking}

In the linear sigma model, the classical ground state is determined by the minimum of the 
potential term in the Lagrangian which describes 
the self--interaction of the scalars 
\begin{equation}
V\left( \sigma ,\bm{\pi}\right) =\frac{\mu ^{2}}{2}
\left( \sigma ^{2}+\bm{\pi }^{2}\right) +\frac{\lambda }
{24}\left( \sigma ^{2}+\bm{\pi}^{2}\right) ^{2}~.
\end{equation}
If the mass term  $\mu^2$ is negative there is a  minimum of the potential which is located at 
\begin{equation}
\sigma ^{2}+\bm{\pi}^{2}=-\frac{6\mu ^{2}}{\lambda} \equiv v^{2}~.
\end{equation}
This  defines  a 3--sphere, $S^{3}$, in the 4--dimensional space formed by the scalar
fields. Each point on $S^{3}$ is invariant under $O(3)$
rotations. For example, the point $(v,0,0,0) $ is invariant
under the rotations of the last three components of the vector. This means that  after a
point on $S^{3}$ is chosen to be the classical ground state, the symmetry is
broken spontaneously from $O(4)$ to $O(3)$ We
should point out that the vector $SU(2)$ isospin symmetry is 
isomorphic to $O(3)$~.

Quantizing the theory  we will need to expand the fields around the classical
values,  however we may choose a particular ground state such as 
\begin{equation}
<\sigma >=v\equiv -\frac{6\mu ^{2}}{\lambda }\qquad
\quad \textrm {and} \qquad <\bm{\pi}>=0~.
\end{equation}
\noindent
The constant quantity $v$ is usually called the vacuum expectation value (VEV). Then we see
that we can expand the sigma field around the minimum as
$\sigma =v+\sigma ^{\prime}$, so the Lagrangian becomes
\begin{eqnarray}
\mathcal{L} &=&\frac{1}{2}\left( \partial _{\mu}\sigma^{\prime} \right)^{2} 
+\frac{1}{2}\left( \partial _{\mu} \bm{\pi}\right)^{2}
+\mu^{2}{\sigma^{\prime}}^{2}\\ \nonumber
&&-\frac{\lambda v}{6}\sigma^{\prime}
\left({\sigma^{\prime}}^{2}+\bm{\pi}^{2}\right)
-\frac{\lambda }{24}\left( {\sigma^{\prime}} ^{2}+\bm{\pi}^{2}\right)^{2}
\\ \nonumber
&&+\bar{\psi}i\gamma ^{\mu}\partial _{\mu}\psi+g v \bar{\psi}\psi
+ig\bar{\psi}\bm{\tau}\gamma_{5}\psi \cdot \bm{\pi} 
+g\bar{\psi}\psi \sigma^{\prime}~.   
\end{eqnarray}

We observe that  the fermions are massive while the pions appear massless. This is a
consequence of the Goldstone theorem. According this theorem  spontaneous symmetry
breaking  of a continuous symmetry will result in massless particles, or zero
energy excitations. 

After SSB, the original multiplet $(\sigma,\bm{\pi})$, splits into 
massless pions and a massive sigma.  Also the fermions acquire mass which
is proportional to VEV.  Therefore, although the interaction is $SU(2)
_{L}\times SU(2) _{R}$ symmetric, the particle spectrum 
is only isospin $SU(2)$ symmetric. This is the typical consequence 
of SSB. In some sense,
the original symmetry is realized by combining the $SU(2)$
multiplet with the massless Goldstone bosons to form the
multiplets of $SU(2) _{L}\times SU(2) _{R}$. The sigma field is massive 
with mass $m^2=-2\mu^2=\lambda v^2/3$ and 
there are trilinear $\sigma$--$\pi$ couplings proportional to $v$. Another 
important point is that after the SSB, the axial current 
will have a term linear in $\pi $ field, 
\begin{equation}
A_{i}^{\mu }=iv\partial _{\mu }\pi _{i}+\cdots
\end{equation}
which is responsible for the matrix element, 
\begin{equation}
<0|A_{i}^{\mu }|\pi _{j}\left( p\right) >=ip^{\mu }v~.
\end{equation}
Using this matrix element in $\pi $ decay, we can associate  the VEV $v$ with
the pion decay constant $f_{\pi }.$ This coupling between axial current 
$A_{i}^{\mu }$ and $\pi _{i}$ causes the appearance of a massless pole.

In the scalar self interaction, quartic, cubic, and
quadratic terms have only 2 parameters, the coupling constant $\lambda$ and 
the mass term $\mu$. This means that
these three terms are not independent, and there is a relation among them. This
is an example of the low energy theorem for a theory with spontaneous symmetry
breaking. We discuss briefly the low energy theorem in Section~\ref{Sect:Low-Energy-Theorem}

\subsubsection{Explicit breaking of chiral symmetry}

The $SU(2)_L\times SU(2)_R$ symmetry of the linear sigma model 
is explicitly broken if the potential $V(\sigma,\bm{\pi})$ is made 
slightly asymmetric,  e.g.
by the addition of the term
\begin{equation}
\mathcal{L}_{b}=\epsilon \sigma
\end{equation}
to the basic Lagrangian in Eq.~(\ref{Eq:Full-Lagrangian}). To first order
in the quantity $\epsilon$, this shifts the minimum of the potential
to 
\begin{equation}
v=\left(\frac{6\mu ^{2}}{\lambda}\right)^{1/2}+\;\frac{\epsilon}{2\mu^2}~.
\end{equation}
\noindent
As a result the pions acquire mass given by
\begin{equation}
m_\pi^2=\frac{\epsilon}{v}~.
\end{equation}

\subsubsection{\label{Sect:Low-Energy-Theorem}Low energy theorem}

This theorem is one of the  most distinctive  relations among amplitudes
involving Goldstone bosons at low energies. These relations are a consequence
of the fact that Goldstone bosons are
massless. Since Goldstone bosons do carry energies, this is possible only 
to the limit that Goldstone bosons have zero energies. Alternatively 
we can think of zero--energy Goldstone bosons as infinitesimal chiral 
rotation of vacuum. In a chirally  symmetric theory, it has no effect.
  
\begin{figure}
\includegraphics[scale=0.53]{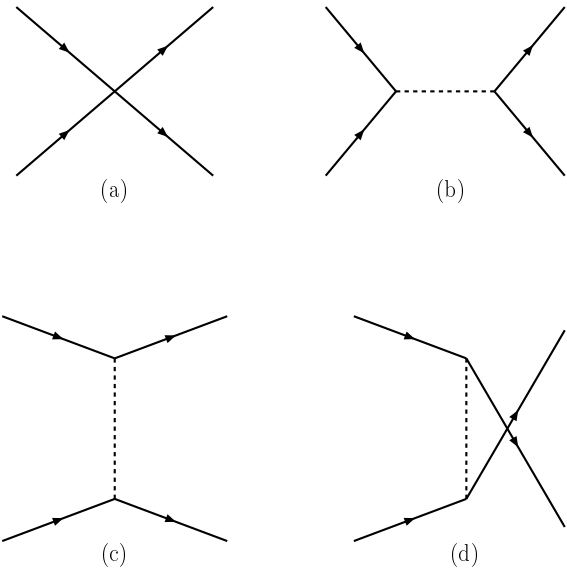} 

\caption{Tree--level graphs contributing to pion scattering at zero 
temperature. Continuous lines represent pions and dashed lines
correspond to sigma. Time runs from left to right.}
\label{Fig:Tree-Graphs}
\end{figure}

Consider the following process involving pion elastic scattering 
with or without sigma exchange. The tree--level contributions come 
from the diagrams 
in Fig.~\ref{Fig:Tree-Graphs}. The amplitudes for these diagrams are given by, 
\begin{eqnarray}
&& iM_{a}=-i\lambda~, \\ \nonumber 
&& iM_{b}=\left(\frac{-2i\lambda v}{6}\right)^2\frac{i}{s-m_\sigma^2}~,\\ \nonumber
&& iM_{c}=\left(\frac{-2i\lambda v}{6}\right)^2\frac{i}{t-m_\sigma^2}~,\\ \nonumber 
&& iM_{d}=\left(\frac{-2i\lambda v}{6}\right)^2\frac{i}{u-m_\sigma^2}~,
\end{eqnarray}
\noindent
where $s,$ $t,$ and $u$ are the usual Mandelstam 
variables,
\begin{equation} 
s=(p_{1}+p_{2})^{2},\quad t=(p_{1}-p_{3})^{2},\quad u=(p_{1}-p_{4})^{2}~.
\end{equation} 
Then the total amplitude is
\begin{eqnarray}
iM&=&iM_{a}+iM_{b}+iM_{c}+iM_{d}\\ \nonumber
&=&-i\lambda-\frac{i\lambda^2 v^2}{9}\left(\frac{1}{s-m_\sigma^2}
+\frac{1}{t-m_\sigma^2}+\frac{1}{u-m_\sigma^2}\right)~.
\end{eqnarray}
\noindent
In  the limit where pions have zero momenta,  we get 
$s,t,u \longrightarrow 0$, and the amplitude is
\begin{equation}
iM=-i\lambda+\frac{i\lambda^2 v^2}{9}\frac{3}{m_\sigma^2}~.
\end{equation}
For our choice of parameters of  the linear 
sigma model, $3m_\sigma^2=\lambda v^2$. Thus, the
amplitude vanishes in the soft pion limit.

This is equivalent to taking  the limit, $m_{\sigma}{^2}\rightarrow \infty$. Soft pion means 
that the pion momentum is much smaller than the sigma mass $m_\sigma$. These are
simple examples of the low energy theorem which states that physical
amplitudes vanish at the limit where the mass of Goldstone 
bosons goes to zero.

\subsection{\label{Sect:Potential}The thermal effective potential}

\subsubsection{The conventional effective potential}

\begin{figure}
\includegraphics[scale=0.53]{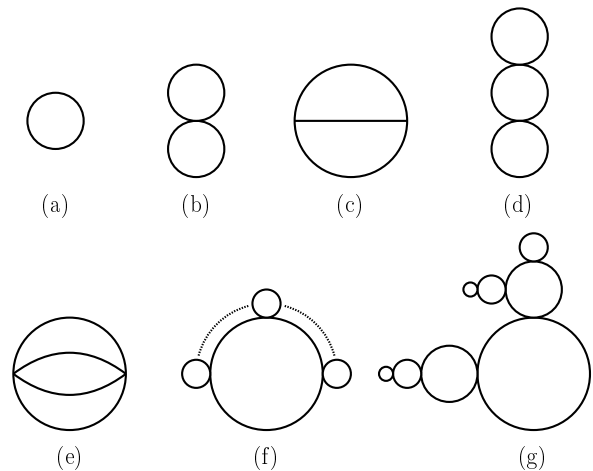}
 
\caption{Examples  of the various types of 1PI diagrams which contribute
to the effective potential for a $\lambda \phi^4$ theory. In 
particular, diagrams of the type (f), are called ``daisies'' and those 
of type (g), super--daisies, or foam  diagrams respectively.}
\label{Fig:1PI}  
\end{figure}

The finite--temperature effective potential  
$V(\phi,T)$,  is defined through an effective  
action $\Gamma (\phi)$,  which is the  
generating functional  of the one particle irreducible graphs 
(a graph is called one--particle irreducible (1PI) if it cannot become 
disconnected by opening only one line, otherwise it is one--particle 
reducible), having the meaning of  the free energy of the 
system. Diagrammatically, the loop expansion of the effective 
potential could appear as
given in Fig.~\ref{Fig:1PI}. Calculations using the loop expansion are 
very difficult beyond  
two loops. One way out of this problem is to perform selective summations 
of higher--loop graphs.

One way to systematise such summations is the large--$N$ method, in which
one considers an $N$--component field, and uses the fact that in
the large $N$ limit, some multi--loop graphs give
greater contributions than others.
For example in the $\lambda \Phi^4$, $N$--component scalar theory, the leading
multi--loop contributions come from daisy and superdaisy graphs in
Fig.~\ref{Fig:1PI}. 

\subsubsection {The Cornwall--Jackiw--Tomboulis method}

Another way to perform systematic selective summations
is to use the method of the effective action for composite
operators \cite{Cornwall:1974vz}.  In this case, the effective action is the generating  
functional of the two--particle 
irreducible (2PI) vacuum graphs (a graph is called 
``two--particle irreducible'' if it does not become disconnected 
upon opening two lines). Now, the effective action  $\Gamma(\phi,G)$, depends not only on  
$\phi(x)$, but  on $G(x,y)$, as well. These two quantities are to be realized  
as the possible expectation values of a quantum field  $\Phi(x)$ and as the
time ordered product of the field operator $T\Phi(x)\Phi(y)$  
respectively. This formalism was initially used for a study of the $O(N)$ model at zero 
temperature \cite{Cornwall:1974vz}, but it has been extended to finite--temperature  
by Amelino--Camelia and Pi, and was used  
for investigations of the effective potential of the  
$\lambda\phi^4$ theory \cite{Amelino-Camelia:1993nc}, and gauge  
theories \cite{Amelino-Camelia:1994kd}. 

\begin{figure}
\includegraphics[scale=0.53]{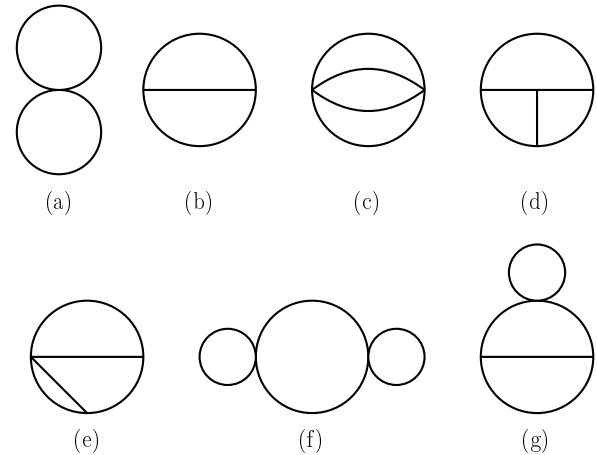} 

\caption{a--e: Examples of  two--particle irreducible graphs which
contribute to the effective potential in the 
CJT method, up to three loops. The lines represent ``dressed''
propagators $G(x,y)$. f--g: These graphs are two--particle reducible. They
do not contribute in the CJT method, however they do contribute to ordinary
effective action $\Gamma(\phi)$. In this case the lines represent the 
tree--level propagator $D(\phi;x,y)$. }
\label{Fig:2PI} 
\end{figure} 

There is an advantage in using the CJT method to calculate the  
effective potential in certain approximations as is, for example, the  
Hartree approximation of the $\lambda\phi^4$ theory.  According to reference 
\cite{Amelino-Camelia:1993nc}, if we use an  
ansatz for a  ``dressed propagator'', we need to evaluate only 
one graph, that of the ``double bubble'' 
in Fig.~\ref{Fig:2PI}a, instead of summing  the infinite class of ``daisy'' and  
``super--daisy'' graphs, given in Figs.\ref{Fig:1PI}f,g, using  the usual  
tree level propagators. 

We demonstrate 
the advantage of this method in Section~\ref{Sect:F4},  where we 
calculate the finite--temperature effective potential for one  
scalar field, with quartic self--interaction. The calculation of 
the effective potential  
by using the CJT formalism is reviewed, in  detail, in 
references \cite{Amelino-Camelia:1993nc,Amelino-Camelia:1996hw}, but 
in order to illustrate 
the method for the calculation of the effective  
potential for the linear sigma model at finite--temperature, we  
will reproduce the basic steps here. For this presentation we follow
closely the derivation as in \cite{Cornwall:1974vz}.

In order to define the effective action for composite
operators, we can  follow a path analogous to the one leading to the
ordinary effective action. The essential difference is that 
the  partition function  depends also on a
bilocal source $K(x,y)$, in addition to the local source $J(x)$.
As an example,  we  consider the $\lambda\Phi^4$ theory with Lagrangian
\begin{equation}
\label{Lagran} 
\mathcal{L}(\Phi) = \frac{1}{2} \, \partial_\mu \Phi \, \partial^\mu \Phi
- \frac{1}{2} \, m^2 \Phi^2 - \frac{\lambda}{24} \, \Phi^4~.
\end{equation} 
According to CJT method \cite{Cornwall:1974vz}, the generating functional 
for the Green functions in the presence
of sources $J(x)$ and $K(x,y)$ is given by (we set $\hbar=1$) 
\begin{equation} 
\mathcal{Z}(J,K) = e^{\mathcal{W}(J,K)} =  \int \mathcal{D}\Phi \exp
      \bigg [I (\Phi) +\Phi  J + \frac{1}{2} \Phi  K \Phi \bigg]
\end{equation}
where $\mathcal{W}(J,K)$, is the generating functional for the
connected Green functions, while  $I(\Phi) = \int d x ^4\, \mathcal{L}(x)$ is
the classical action. Adopting  a similar  notation 
as Rischke and Lenaghan \cite{Lenaghan:2000si}, we have used the shorthands
\begin{eqnarray}
\Phi \, J & \equiv & \int d^4x  \,\Phi(x) \, J(x)~, \\
\Phi\, K \, \Phi & \equiv & 
\int d^4x d^4 y \,\Phi(x) \, K(x,y)\,  \Phi(y) ~.
\end{eqnarray}
The expectation values for the one--point function, $\phi(x)$, in the presence of
a source $J(x)$ is given by
\begin{equation} 
\frac{\delta \mathcal{W}(J,K)}{\delta J(x)} \equiv \phi(x) ~,
\end{equation}
while the connected two--point function, $G(x,y)$ is
\begin{equation}  
\frac{\delta \mathcal{W}(J,K)}{\delta K(x,y)}  \equiv 
        \frac{1}{2} \big [ G(x,y)+ \phi(x) \, \phi(y) \big ]~.
\end{equation} 

The effective action for composite operators 
$\Gamma(\phi,G)$, is obtained  through a double
Legendre transformation  of $\mathcal{W}(J,K)\equiv \ln\mathcal{Z}(J,K)$ 
\begin{equation}
\Gamma(\phi,G) = \mathcal{W}(J,K) - \phi \, J - \frac{1}{2} \, \phi\, K\, \phi -  \frac{1}{2}\, G\, K ~,
\end{equation}
where $G\, K = \int d^4 x d^4 y \, G(x,y)\, K(y,x)$. 
Then  it follows that 
\begin{eqnarray}
\frac{\delta \Gamma(\phi,G)}{\delta \phi(x)} & = & -J(x) - \int d^4 y \, K(x,y)\, \phi(y) ~,\\   
\frac{\delta \Gamma(\phi,G)}{\delta G(x,y)} & = & -\frac{1}{2}\,K(x,y)~. 
\end{eqnarray}

Physical processes correspond to vanishing sources, so the stationarity conditions which determine
the expectation value of the field $\varphi(x)$, and
the (dressed) propagator $\mathcal{G}(x,y)$ are given by  
\begin{eqnarray} 
\frac{\delta \Gamma(\phi,G)}{\delta \phi(x)} \, 
\bigg|_{\phi=\varphi,\, G= \mathcal{G}} &  = & 0~, \\
 \frac{\delta \Gamma(\phi,G)}{\delta G(x,y)} 
\bigg|_{\phi=\varphi,\, G= \mathcal{G}} &  = & 0~. 
\end{eqnarray}

As it was shown by Cornwall--Jackiw--Tomboulis in \cite{Cornwall:1974vz}, the 
effective action $\Gamma(\phi,G)$ 
is given by
\begin{eqnarray} 
\Gamma(\phi,G) &=& I(\phi) - \frac{1}{2} \textrm{Tr}\left(\ln \,G^{-1}\right)\\ \nonumber 
&&- \frac{1}{2}\textrm{Tr}\left(D^{-1}\, G-1\right) + \Gamma_2(\phi,G)~, 
\end{eqnarray}
where in this last equation  $\Gamma_2(\phi,G)$ is the sum of all
two particle irreducible (2PI) diagrams in which all lines represent full (dressed) 
propagators $G$,  while $D^{-1}$ is the inverse of the tree--level propagator,
\begin{equation}
D^{-1} (x,y;\phi) \equiv -   \frac{\delta^2 I(\phi)}{\delta 
\phi(x) \,
\delta \phi(y) } \bigg|_{\phi=\varphi} ~.
\end{equation}

In the case when translation invariance is not broken, it can be assumed  that 
the field $\phi(x) =\textrm{constant}= \phi$. Then an infinite volume factor $\Omega$
arises from space--time integrations and it is customary to introduce the generalized 
effective potential $V(\phi,G)=\Gamma(\phi,G)/\Omega$.  The exact expression for the effective potential 
is then
\begin{eqnarray}
V(\phi,G) &=& U(\phi) +\frac{1}{2} \int_k \, 
\ln G^{-1}(k) \\ \nonumber
&&+ \frac{1}{2}\,\int_k\, \big [D^{-1}(k;\phi)
\, G(k) - 1 \big ] + V_2(\phi,G)~,
\label{Eq:Veff}
\end{eqnarray}
where is $U(\phi)$ is the classical potential of the Lagrangian, $D^{-1}(k;\phi)$ is the inverse
tree level propagator, which for $\lambda\phi^4$ theory is given by
\begin{equation} 
D^{-1}(k;\phi) = k^2 +m^2+\frac{1}{2}\lambda\phi^2~, 
\end{equation}
and $V_2(\phi,G)=\Gamma_2(\phi,G)/\Omega$.
The stationarity conditions for the effective action reduce to the following ones
involving the effective potential
  
\begin{eqnarray} 
 \frac{\delta V(\phi,G)}{\delta \phi} \, 
\bigg|_{\phi=\varphi,\, G= \mathcal{G}}= 0 ~,\\
 \frac{\delta V(\phi,G)}{\delta G(k)} \,
\bigg|_{\phi=\varphi,\, G= \mathcal{G}}= 0~. 
\end{eqnarray}
We then obtain the following relation 
\begin{equation}
\mathcal{G}^{-1}(k) = D^{-1}(k;\varphi) + \Sigma(k)~ ,
\end{equation}
\noindent 
which corresponds to a Schwinger--Dyson equation for the full (dressed) propagator $\mathcal{G}$
and where the self--energy $\Sigma(k)$ is given by
\begin{equation}
\Sigma(k) \equiv 2   \frac{\delta V_2 (\phi,G)}{\delta G(k)} 
\bigg|_{\phi=\varphi,\, G= \mathcal{G}}~. 
\end{equation}

The thermal effective potential has the meaning of the free energy density and it is related 
to thermodynamic pressure through the equation
\begin{equation}
p = - V (\varphi,\mathcal{G})~.
\end{equation}

\subsection{\label{Sect:Formalisms}Finite temperature formalisms}

Whatever the method  one chooses to calculate the effective 
potential, the calculation has to be performed in the framework of finite
temperature field theory.  There are three main paths to 
follow namely: thermo--field dynamics, imaginary
time formalism and real time formalism. In our calculations, we do 
not use the machinery of thermo--field dynamics, but we use 
both real and imaginary time. Each of these has certain advantages and 
disadvantages. Traditionally, the imaginary time seems to be convenient for 
systems in equilibrium, since the real time is thought to be the 
appropriate for 
calculations 
in systems far from equilibrium.

\subsubsection{Imaginary time formalism}
The imaginary time formalism, which is  
also known as the  
Matsubara formalism, provides a way of evaluating
the partition function perturbatively using a diagrammatic 
method which is analogous to that which is used in conventional 
field theory at zero 
temperature 
\cite{Dolan:1974qd,Kapusta:1989,Das:1997gg,Landsman:1987uw}. According to  
this technique, we work in Euclidean space--time and 
use the  same Feynman rules as at zero--temperature, but  
when evaluating momentum space integrals, we replace integration 
over the time component $k_{4}$ with a summation over discrete 
frequencies. That means that  in the case of  
bosons  $k_{4}=2\pi i n T,\;\;n=0,\pm 1,\pm 2,\pm 3,\ldots$. 
This is encoded into the following relationship 
\begin{equation} 
\int\frac{d^{4}k}{(2\pi)^{4}} f(k)\longrightarrow 
 \frac{1}{\beta}\sum_{n}\int\frac{d^{3}\textbf{k}}{(2\pi)^{3}}  
f(2\pi i n T,\textbf{k}) 
\end{equation} 
where $\beta$ is the inverse temperature, $\beta=1/k_{B}T$, and 
as usual Boltzmann's constant is taken to be $k_{B}=1$. For the sake of  
simplicity, in  
the following we  introduc a subscript 
$\beta$, to denote integration and summation over the Matsubara 
frequency sums. So in what follows we adopt the shorthand expression 
\begin{equation} 
 \frac{1}{\beta}\sum_{n}\int\frac{d^{3}\textbf{k}}{(2\pi)^{3}}  
f(2\pi i n T,\textbf{k}) 
\equiv\int_{\beta}f(2\pi i n T,\textbf{k})~ 
\end{equation} 

If one is to study systems at equilibrium, the imaginary time 
formalism is adequate, but for dynamical systems far from 
equilibrium, we need to analytically continue back to real time 
which was traded in favour of the temperature. An alternative 
way is to work directly   in the
real time formalism from the outset. We introduce the basic concepts about 
the real time formalism in the next section. 
 
\subsubsection{Real time formalism}

This formalism has a long history of applications in condensed 
matter systems, where  early papers appeared in the 1950's. It is also known
as Keldysh or closed--path formalism \cite{Keldysh:ud}. It has only been
quite recently that researchers realized  its usefulness in applications 
in particle physics at finite--temperature. The formalism 
is given  in details in 
\cite{Niemi:1984nf,Niemi:1984ea}, and is also 
reviewed in \cite{Landsman:1987uw}, where a formal account 
for  both -- real time and imaginary time -- formalisms is given. 
There is also a lot of information  in the recent books 
\cite{Rivers:1987,LeBellac:1996,Das:1997gg}. Further information
can also be found in  recent papers \cite{Das:2000ft,Landshoff:1998ku,Altherr:1993tn}. 
 
The basic point is that for a  scalar field theory, the 
propagator becomes a two by two matrix and is divided 
into two parts as 
\begin{equation} 
i\mathcal{D}= i\mathcal{D}_0+i\mathcal{D_\beta}~,
\end{equation} 
where the zero temperature 
part appears as   
\begin{equation} 
i\mathcal{D}_0=
\begin{bmatrix}\displaystyle\frac{i}{k^2-m^2+i\epsilon} & 0 \\ 
             0   & \displaystyle\frac{-i}{k^2-m^2-i\epsilon} \\
\end{bmatrix} 
\end{equation} 
and  the temperature dependent part is of the form 
\begin{equation} 
i\mathcal{D_\beta}=\frac{\epsilon}{(k^2-m^2)^2+\epsilon^2} 
\begin{bmatrix}2\sinh^2\theta & \sinh\;2\theta \\ 
             \sinh\;2\theta   & 2\sinh^2\theta \\
\end{bmatrix}~, 
\end{equation} 
where 
\begin{equation} 
\cosh^2 \theta = \frac{{\rm e}^{\beta |k_0|}}{{\rm e}^{\beta |k_0|}-1}~. 
\end{equation} 
In the limit $\epsilon \rightarrow 0$, the 
pre--factor of the thermal part reduces to a delta function.

In particular the 
$(1,1)$ component is given by 
\begin{equation} 
\mathcal{D}^{11}=\frac{i}{k^2-m^2+i\epsilon}+2\pi\delta(k^2-m^2) n(k_0) 
\end{equation} 
where $n(k_0)$ is the Bose--Einstein distribution function 
\begin{equation} 
n(k_0)=\frac{1}{\exp(\beta |k_0|)-1}~. 
\end{equation} 
In this last equation $\beta=1/T$ is the inverse temperature,   
while  $k_0=\sqrt{\textbf{k}^2+m^2}$.

\subsection{\label{Sect:F4}The $\lambda\Phi^{4}$ theory at finite temperature}

As an illustration, we use the CJT formalism to calculate the 
effective potential for a $\lambda\Phi^{4}$ theory. The Lagrangian
is given by
\begin{equation} 
\mathcal{L} = \frac{1}{2} (\partial_{\mu} \Phi) (\partial^{\mu} \Phi) 
-\frac{1}{2} m^2 \Phi^2 
-\frac{1}{24}\lambda \Phi^4 ~, 
\label{lf} 
\end{equation}and in 
order to realize the spontaneous breaking of symmetry, $m^2$ 
is considered as a negative parameter. By  
shifting the field as $\Phi\rightarrow \Phi+\phi$, the  
``classical potential'' takes the form  
\begin{equation} 
U(\phi)=\frac{1}{2} m^2 \phi^2 
+\frac{1}{24}\lambda \phi^4~, 
\label{classical} 
\end{equation} 
and the interaction Lagrangian which  
describes the vertices of the shifted, theory is given by 
\begin{equation} 
\mathcal{L}_{int} = 
-\frac{\lambda}{6} \phi\Phi^3 - \frac{\lambda}{24} \Phi^4 ~. 
\end{equation} 
The tree--level propagator which corresponds to the  
above Lagrangian density is 
\begin{equation} 
\mathcal{D}^{-1}(\phi;k) = k^2 +m^2 + \frac{1}{2}\lambda\phi^2  ~. 
\end{equation} 
 
\begin{figure}
\includegraphics[scale=0.5]{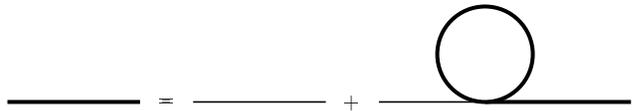} 

\caption{Schematic representation of the self--energy contribution  to scalar
propagator in the self--consistent Hartree approximation. The full or 
``dressed'' propagator appears as a sum of the bare propagator, and the 
self--energy insertion.}
\label{Fig:14}
\end{figure}

According to CJT formalism \cite{Amelino-Camelia:1993nc}, the 
finite temperature effective potential is given by  
\begin{eqnarray}   
V(\phi,G) &=& U(\phi)+\frac{1}{2}\int_{\beta}\ln  
              G^{-1}(\phi;k)\nonumber\\ 
          & &+\frac{1}{2}\int_{\beta} 
              [\mathcal{D}^{-1}(\phi;k)G(\phi;k)-1]\nonumber\\ 
          & &+V_2(\phi,G)~, 
\end{eqnarray} 
where $U(\phi)$ is the ``classical potential'' given by 
Eq.~(\ref{classical}), and $V_2(\phi,G)$ represents 
the infinite sum of the  
two--particle irreducible vacuum graphs. We are  
going to evaluate the effective potential 
in the Hartree approximation which means that we only need 
to calculate the ``double bubble'' 
diagram given in Fig.~\ref{Fig:2PI}\,(a). The resulting  
effective  
potential, is therefore, 
\begin{eqnarray} 
V(\phi,G) &=& \frac{1}{2}m^2 \phi^2  
+\frac{1}{24}\lambda \phi^4 
+\frac{1}{2} \int_{\beta} \ln G^{-1}(\phi;k)  
\nonumber\\ 
& & +\frac{1}{2}\int_{\beta}  
[(k^2 + m^2 + \frac{1}{2}\lambda  \phi^2)G(\phi;k) -1] \nonumber\\   
& & + \frac{1}{8}\lambda  
\big [ \int_{\beta}  G(\phi;k) \big ]^2 ~.  
\label{potential} 
\end{eqnarray} 
Minimizing the effective potential with respect 
to the ``dressed propagator'' $G(\phi;k)$, we obtain the gap equation 
\begin{equation} 
G^{-1}( \phi;k) = k^2 + m^2 + \frac{1}{2}\lambda \phi^2 
+ \frac{1}{2}\lambda\int_{\beta} G(\phi;k) ~.  
\end{equation}
In graphical terms, we could 
say that
the self--energy is calculated by opening one line of each diagram contained 
in the functional of the effective potential 
\cite{vanHees:2000bp,Ivanov:2000ma}. In Hartree
approximation, this is illustrated 
in  the  Fig.~\ref{Fig:14}~.

The solution $G_0(\phi;k)$,  of the gap equation, is inserted back 
into the expression for the effective potential, resulting in a potential
which is  a function of $\phi$. As is stated in 
\cite{Cornwall:1974vz}, using  
the propagator $G_{0}(\phi;k)$ for internal lines  
corresponds to summing all daisy and super-daisy 
diagrams, using the usual tree level propagators as in \cite{Dolan:1974qd}.

Then,  we can
adopt the following form for the dressed
propagator $G(\phi;k)$, 
\begin{equation} 
G(\phi;k)= \frac{1}{k^2 + M^2}~, 
\end{equation} 
where an ``effective mass'' $M=M(\phi;k)$, has been  
introduced. The gap equation for the propagator then  becomes an equation 
for the effective mass 
\begin{equation} 
M^2 = m^2+\frac{\lambda}{2}\phi^2+  
\frac{\lambda}{2}\int_{\beta}\frac{1}{k^2 + M^2 }~, 
\label{Eqn:gap} 
\end{equation} 
where it is obvious, since we integrate over the loop momenta, that in 
this approximation, the effective mass $M$ is momentum independent.

In terms of the solution $M_{0}(\phi)$ of  the gap  
Eq.~(\ref{Eqn:gap}), the 
effective potential takes the form 
\begin{eqnarray} 
V( \phi,M_{0}) &=& \frac{1}{2} m^2 \phi^2 
+\frac{\lambda}{24}\phi^4+\frac{1}{2}\int_{\beta}\ln  
              (k^{2}+M^{2}_{0}) \nonumber \\ 
          & & -\frac{1}{2}(M^2_{0} - m^2 - \frac{\lambda}{2} \phi^2 ) 
\int_{\beta}\frac{1}{k^2 + M^2_{0}} \nonumber \\
          & & + \frac{\lambda}{8}\big [ \int_{\beta} 
              \frac{1}{k^2 + M_{0}^2}\big ]^{2}~.  
\end{eqnarray}

Performing the Matsubara frequency sums as in \cite{Dolan:1974qd}
(we repeat these steps in appendix A), the  
logarithmic integral which appears in  
the above expression for the effective potential 
divides into two parts. A zero--temperature part, $Q_{0}(M)$, which 
is divergent, and a nonzero--temperature part, $Q_{\beta}(M)$,  which 
is finite, and can be written as 
\begin{eqnarray} 
Q(M)&=&\frac{1}{2}\int_{\beta}\ln(k^{2}+M^{2})\nonumber\\ 
&=&Q_{0}(M)+Q_{\beta}(M) \nonumber\\ 
&=&\int\frac{d^3 \textbf{k}}{(2\pi)^3}\frac{\omega_{\textbf{k}}}{2}\nonumber\\ 
&&+\frac{1}{\beta}
\int\frac{d^3 \textbf{k}}{(2\pi)^3} 
\ln[1-\exp(-\beta\omega_{\textbf{k}})]~, 
\end{eqnarray} 
where $\omega_{\textbf{k}}=(\textbf{k}^{2}+M^{2})^{1/2}$. In this 
above expression, and in what follows, we omit the subscript 0 on $M$. 
Similarly, the second integral is divided into a 
zero--temperature part $F_{0}(M)$, and a finite--temperature  
part $F_{\beta}(M)$, given by  
\begin{eqnarray} 
F(M)&=&\int_{\beta} \frac{1}{k^2 + M^2 }\nonumber\\  
&=&F_{0}(M)+F_{\beta}(M)\nonumber\\ 
&=&\int\frac{d^3 \textbf{k}}{(2\pi)^3}\frac{1}{2\omega_{\textbf{k}}}\nonumber\\  
&&+\int\frac{d^3 \textbf{k}}{(2\pi)^3}\frac{1}{\omega_{\textbf{k}}} 
\frac{1}{\exp(\beta\omega_{\textbf{k}})-1}~. 
\end{eqnarray} 
The second term vanishes at zero temperature, while  
the first term survives but it gives 
rise to divergences which can be carried out  
using appropriate renormalization prescriptions 
\cite{Amelino-Camelia:1993nc,Amelino-Camelia:1996hw}. 
If one is interested  in temperature induced   
effects only, as is our 
approximation, the divergent integrals can be ignored. In this case, by  
making a change to the 
integration variables, the finite--temperature part of  
$F(M)$ can be written as 
\begin{equation} 
F_{\beta}(M)=\frac{T^{2}}{2\pi^{2}}\int_{0}^{\infty}\frac{x^2\, dx} 
{(x^{2}+y^{2})^{1/2}} 
\frac{1}{\exp(x^{2}+y^{2})^{1/2}-1}~, 
\label{Eq:FM} 
\end{equation} 
where we have used a shorthand notation, $y=M/T$. 

In order to simplify the
lengthy expressions involved in what follows, we
introduce the function $g(z,a)$ defined as
\begin{equation}
g(z,a)=\frac{1}{(z^2+a^2)^{1/2}}
\frac{1}{\exp(z^2+a^2)^{1/2}-1}~.
\label{Eq:g(z,a)}
\end{equation}
Then the function $f(a)$, will appear as the shorthand of
the integral
\begin{eqnarray}
f(a)&=&\int_0^\infty\frac{x^2 dx}
{(x^2+a^2)^{1/2}}
\frac{1}{\exp(x^2+a^2)^{1/2}-1} \\ \nonumber
&=&\int_0^\infty x^2\, dx \, g(x,a)~,
\end{eqnarray}
and in the limit of $a=0$, we recover the well known
result $f(0)=\pi^2/6$.

Similarly, the finite--temperature part of the logarithmic integral 
becomes 
\begin{equation} 
Q_{\beta}(M)=\frac{T^4}{2\pi^2}\int_{0}^{\infty}x^2 \,dx  
\ln\left (1-\exp[-(x^2+y^2)^{1/2}]\right )~. 
\label{Eq:QM} 
\end{equation}  
Then, the finite--temperature effective potential, ignoring quantum  
fluctuations, can be written as  
\begin{eqnarray} 
V(\phi,M)&=&\frac{1}{2}m^{2}\phi^{2}+\frac{1}{24}\lambda\phi^{4} 
+ Q_{\beta}(M) \nonumber\\ 
&&\mbox{}-\frac{1}{2}(M^{2}-m^{2}-\frac{\lambda}{2}\phi^{2})F_{\beta}(M) \nonumber\\  
&&+\frac{\lambda}{8}[F_{\beta}(M)]^{2}~. 
\end{eqnarray} 
We can obtain a more compact form if we make use of the gap equation  
(\ref{Eqn:gap}),  
\begin{equation} 
V(\phi,M)=\frac{1}{2}m^{2} \phi^{2}+\frac{1}{24}\lambda \phi^{4} 
+ Q_{\beta}(M)-\frac{1}{8}\lambda\big [F_{\beta}(M)\big ]^{2}~. 
\end{equation}

\section{\label{Sect:Hartree}Hartree approximation}   

\subsection{Introduction} 

In this section, we apply the CJT
method in order to calculate the effective potential 
of the $O(4)$ linear sigma model. The $\sigma$ field can be used to represent 
the quark condensate, the order parameter for 
the chiral phase transition, since both exhibit the same behaviour 
under chiral transformations \cite{Birse:1994cz,Hatsuda:2000by,Kunihiro:2000je}. The pions  
are very light particles, and can 
be considered  approximately as massless Goldstone  
bosons. We use the model as an effective 
theory for QCD, ignoring 
the fermion sector for the moment, and  
concentrating on the meson  
sector. We examine this model 
in two approximation schemes: (a) in the chiral limit
considering the pions true Goldstone bosons, and therefore
massless, and (b) treating them as massive. 

\subsection{The linear sigma model} 

As we mentioned in the introduction, the linear sigma model 
serves as a good low energy effective theory  
in order for one to gain some insight into QCD. Recently, it has 
attracted much attention, especially in studies involving disoriented chiral 
condensates (DCC's) \cite{Bjorken:1992xr,Abada:1998mb,Randrup:1997es,Randrup:1996ay,Rajagopal:1993ah}.
The model is very well suited for  
describing the physics of pions in studies 
of chiral symmetry. Fermions may be included in the model either 
as nucleons, if one is to study nucleon  
interactions, or as quarks. The mesonic part of the model  
consists of four scalar fields, one 
scalar isoscalar field which is called the sigma $\sigma$ field  and  
the usual three pion fields $\pi^{0}$, $\pi^{\pm}$, which form a  
pseudo--scalar isovector. The fields form a  
four vector $(\sigma,\pi_{i}),\;\;i=1,2,3$, which we regard as the 
chiral field and the model 
displays an $O(4)$ symmetry. 
 
The meson sector Lagrangian of the $O(4)$ sigma model is 
\begin{eqnarray} 
\mathcal{ L}&=&\frac{1}{2}(\partial\sigma)^2 + \frac{1}{2}(\partial \bm{\pi})^2 
        -\frac{1}{2}m^{2}\sigma^2 
-\frac{1}{2}m^{2} 
 \bm{\pi}^2 \nonumber \\
&&- \frac{\lambda}{24}(\sigma^2 + 
 \bm{\pi}^2)^2 - \varepsilon \sigma~.
\label{Eq:Lagrangian-4} 
\end{eqnarray} 
The last term, $\varepsilon \sigma$, has been introduced into 
the above expression   in order to generate masses 
for the pions.  It is related to pion mass as
$\varepsilon=f_{\pi}m_{\pi}^{2}$, where  $f_{\pi}=93\,\,\rm MeV$, is the 
pion decay constant. In the absence of the last term 
which breaks the chiral symmetry, the pions are massless.

The coupling  
constant $\lambda$ of the model can be related to zero temperature 
properties of the pions, and sigma through the expression 
\begin{equation} 
\lambda=\frac{3(m_{\sigma}^{2}-m_{\pi}^{2})}{f_{\pi}^{2}}~. 
\end{equation} 
The negative mass parameter $m^{2}$, is introduced in order 
to obtain spontaneous breaking of symmetry and its value is chosen to be 
\begin{equation} 
-m^{2}=(m_{\sigma}^{2}-3 m_{\pi}^{2})/2 > 0~. 
\end{equation}
To define the numerical values of $\lambda$ and $m^2$, we use
$m_{\pi}\approx 138\,\textrm{MeV}$, since for the  sigma mass 
we adopt  $m_{\sigma}\approx 600\,\textrm{MeV}$.

\subsection{The chiral limit $\varepsilon=0$} 
 
In order to deal with the exact chiral limit first, the starting point 
is the Lagrangian given in Eq.~(\ref{Eq:Lagrangian-4}), and we  ignore 
the symmetry breaking term for the moment.  The parameters
$\lambda$ and $m^2$ 
appearing in the Lagrangian are simply
\begin{equation} 
\lambda=\frac{3m_{\sigma}^{2}}{f_{\pi}^{2}},\qquad 
-2m^{2}=m_{\sigma}^{2} > 0~. 
\end{equation} 
By shifting the sigma field as 
$\sigma\rightarrow \sigma+\phi$, the resulting ``classical potential'' is 
\begin{equation} 
U(\phi)=\frac{1}{2}m^{2}\phi^{2}+\frac{\lambda}{24}\phi^{4}~, 
\end{equation}          
and the interaction   
Lagrangian which  
describes the vertices of the new  
theory takes the form 
\begin{equation} 
\mathcal{L}_{int} = 
- \frac{\lambda}{24} \sigma^4  
-\frac{\lambda}{24}{\bm{\pi}}^4 
-\frac{\lambda}{12}\sigma^2{\bm{\pi}}^2 
-\frac{\lambda}{6} \phi\sigma^3  
-\frac{\lambda}{6} \phi\sigma{\bm{\pi}}^{2} ~. 
\label{Eq:Lint} 
\end{equation} 
In the Hartree approximation, we do not consider interactions 
given by the last two terms in the Lagrangian. We attempt to 
include these interactions in Section~\ref{Sect:Beyond}.
 
The tree level sigma, and pion propagators, corresponding to 
the above Lagrangian are
\begin{eqnarray}
\label{Eq:Propagators-4}
\mathcal{D}^{-1}_{\sigma}(\phi;k) &=& k^2 +m^2 + 
\frac{1}{2}\lambda\phi^2 ~,  \\
\mathcal{D}^{-1}_{\pi}(\phi;k) &=& k^2 +m^2 +\frac{1}{6}\lambda\phi^2  ~.
\end{eqnarray}

\begin{figure}
\includegraphics[scale=0.55]{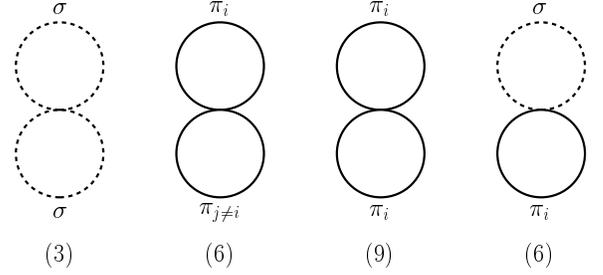} 

\caption{\label{Fig:Bubbles}Graphs of the ``double--bubble'' type which contribute to 
the effective potential for the O(4) linear sigma model in  
the Hartree approximation. The numbers show the weight 
of each type of bubble, in the expression for the effective 
potential.}
\end{figure}

We evaluate the effective potential in the Hartree 
approximation, which means that we only need  to calculate  
the ``double bubble'' 
diagrams as in $\lambda\phi^{4}$ theory. In the linear sigma  
model, the corresponding  
effective potential at finite temperature  can be written as 
\begin{eqnarray} 
V(\phi,G)&=&U(\phi)+\frac{1}{2}\int_{\beta}\ln  
              G^{-1}_{\sigma}(\phi;k) 
+\frac{3}{2}\int_{\beta}\ln G^{-1}_{\pi}(\phi;k) 
\nonumber\\ 
          & &+\frac{1}{2}\int_{\beta} 
              [\mathcal{ D}_{\sigma}^{-1}(\phi;k)G_{\sigma}(\phi;k)-1]
\nonumber\\  
          & &+\frac{3}{2}\int_{\beta} 
              [\mathcal{ D}_{\pi}^{-1}(\phi;k)G_{\pi}(\phi;k)-3]\nonumber\\ 
          & &+V_2(\phi,G_{\sigma},G_{\pi})~, 
\end{eqnarray} 
where the first term $U(\phi)$, is the classical potential
and the last term $V_2(\phi,G_{\sigma},G_{\pi})$, originates from the 
sum of ``double bubble'' diagrams. There 
are four types 
of double bubbles as we show in Fig.~\ref{Fig:Bubbles}, and these contribute  
the following terms in the potential 
\begin{eqnarray} 
V_2(\phi,G_{\sigma},G_{\pi})&=&3\frac{\lambda}{24}\left [ 
\int_{\beta}G_{\sigma}(\phi;k)\right ]^{2} \nonumber \\  
& &+6\frac{\lambda}{24}\int_{\beta}G_{\sigma}(\phi;k) 
\int_{\beta}G_{\pi}(\phi;k)\nonumber \\  
& &+15\frac{\lambda}{24}\left [ 
\int_{\beta}G_{\pi}(\phi;k)\right ]^{2}~. 
\end{eqnarray}

Minimizing the effective potential with respect  
to the ``dressed'' propagators,  we get the following pair  
of nonlinear gap equations 
\begin{eqnarray*}
G_{\sigma}^{-1}(\phi;k)&=&\mathcal{ D}_{\sigma}^{-1}+\frac{\lambda}{2} 
\int_{\beta}G_{\pi}(\phi;k) 
+\frac{\lambda}{2}\int_{\beta}G_{\sigma}(\phi;k)~,
\nonumber\\
G_{\pi}^{-1}(\phi;k)&=&\mathcal{D}_{\pi}^{-1}+\frac{5\lambda}{6} 
\int_{\beta}G_{\pi}(\phi;k) 
+\frac{\lambda}{6} 
\int_{\beta}G_{\sigma}(\phi;k)~.
\nonumber\\ 
\label{Eq:gap-prop} 
\end{eqnarray*} 
The bare propagators $\mathcal{ D}_{\sigma}$ and $\mathcal{D}_{\pi}$ are given 
by Eq.~(\ref{Eq:Propagators-4}). In order to solve this system, we can use  the same  
ansatz for the dressed propagators as  
in the one field case
\begin{equation} 
G_{\sigma/\pi}^{-1}=k^{2}+M^{2}_{\sigma/\pi}~. 
\label{Eq:Propagators-Dressed}  
\end{equation} 
Then by using  Eqs.~(\ref{Eq:Propagators-4}), (\ref{Eq:gap-prop})
the nonlinear system for the dressed propagators reduces 
to  the following system for the thermal 
effective  masses 
\begin{subequations}
\label{Eq:System-H}  
\begin{eqnarray}
M_{\sigma}^{2}&=&m^{2}+\frac{1}{2}\lambda\phi^{2}+ 
\frac{\lambda}{2}F(M_{\sigma})+\frac{\lambda}{2}F(M_{\pi}) 
\\ \label{Eq:System-H.a}
M_{\pi}^{2}&=&m^{2}+\frac{1}{6}\lambda\phi^{2}+\frac{\lambda}{6}F(M_{\sigma}) 
+\frac{5\lambda}{6}F(M_{\pi})~.
\label{Eq:System-H.b}
\end{eqnarray}
\end{subequations}  
In these last two equations, we have used a  shorthand 
notation and $F(M_{\sigma/\pi})$ is given by 
\begin{equation} 
F(M_{\sigma/\pi})=\int_{\beta}\frac{1}{k^2+M^{2}_{\sigma/\pi}}~. 
\end{equation} 
As in $\lambda\phi^{4}$ theory, the thermal effective masses  are  
independent of momentum and functions of the order 
parameter $\phi$, and the  
temperature $T$. 
 
By using these two equations, the effective potential at finite  temperature  
can be written as 
\begin{eqnarray} 
V(\phi,M)&=&\frac{1}{2}m^{2}\phi^{2}+\frac{1}{24}\lambda\phi^{4} \nonumber \\ 
&& + \frac{1}{2}\int_{\beta}\ln(k^{2}+M_{\sigma}^{2})
+ \frac{3}{2}\int_{\beta}\ln(k^{2}+M_{\pi}^{2}) \nonumber\\ 
&& - \frac{1}{2}(M_{\sigma}^{2}-m^{2} -\frac{1}{2}\lambda\phi^{2})F(M_{\sigma})\nonumber\\  
&& - \frac{3}{2}(M_{\pi}^{2}-m^{2}-\frac{1}{6}\lambda\phi^{2})F(M_{\pi}) \nonumber\\ 
&& + \frac{\lambda}{8}[F(M_{\sigma})]^{2}+\frac{5\lambda}{8} [F(M_{\pi})]^{2}\nonumber\\ 
&& + \frac{\lambda}{4}F(M_{\sigma})F(M_{\pi})~. 
\end{eqnarray} 
Minimizing the effective potential with respect  
to the ``dressed'' propagators,  we  have found   
the set of nonlinear gap equations for the effective 
particles' masses, given by Eq.~(\ref{Eq:System-H}). In addition, by minimizing  
the potential with respect to the order parameter, we obtain one  
more equation 
\begin{equation} 
0=m^{2}+\frac{1}{6}\lambda\phi^{2}+\frac{\lambda}{2}F(M_{\sigma}) 
+\frac{\lambda}{2}F(M_{\pi})~. 
\label{Eq:Phi-H} 
\end{equation} 
In order to study the evolution of the potential as a function of  
temperature, we 
perform the Matsubara frequency sums,  as in  
the one field case. There are some problems 
concerning the renormalization of the model  
\cite{Amelino-Camelia:1993nc,Amelino-Camelia:1997dd,Baym:1977qb,Roh:1998ek}. At 
the level of our 
approximation, we ignore  
quantum fluctuations for the moment, and keep  only the finite  
temperature part of the  
integrals. This decision is justified by the fact that the finite 
terms -- given by Eq.~(\ref{Eq:FM}) and Eq.~(\ref{Eq:QM}) -- correspond 
to interactions between thermallly excited mesons with other mesons
present in the plasma \cite{Rivers:1994rz}. In doing so, we neglect 
the quantum fluctuations of the meson fields and keep only
the thermal fluctuations present in the hot plasma. The same choice
was also adopted in \cite{Roh:1998ek,Larsen:1986ei}. In a very recent 
investigation 
of the same model, and based on the same formalism, Rischke and Lenaghan 
\cite{Lenaghan:2000si} 
have shown that the model is indeed renormalisable. Renormalization 
of the model has also demonstrated by Chiku and Hatsuda 
\cite{Chiku:1998va,Chiku:1998kd}, who on the basis of optimised perturbation 
theory, and the real 
time formalism, have studied the spectral densities and the effective 
masses of the sigma and the pions. 

\begin{figure}
\includegraphics[scale=0.6]{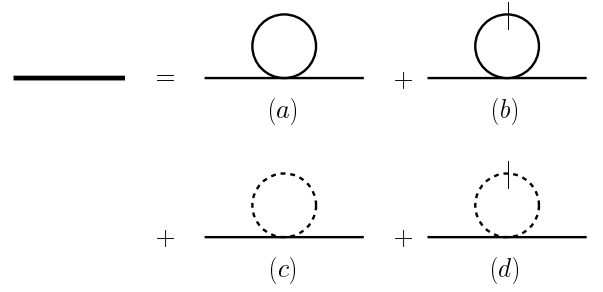} 
\caption{\label{Fig:Pion-Self}Schematic presentation of selfenergy contribution to pion 
propagator. Dashed  lines represent sigma propagator, while continuous ones correspond to pions.}
\end{figure}

\begin{figure}
\includegraphics[scale=0.6]{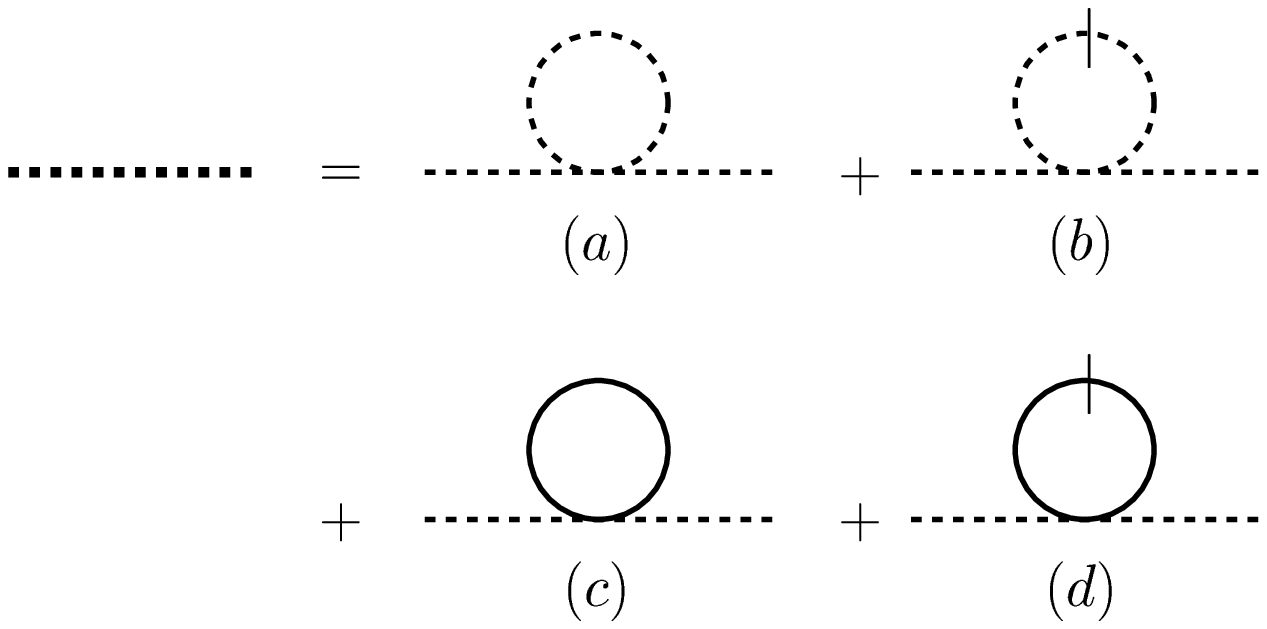} 
\caption{\label{Fig:Sigma-Self}Schematic presentation of selfenergy contribution to sigma 
propagator. Dashed  lines represent sigma propagator, while continuous ones correspond to pions.}
\end{figure}

Diagrammatically, differentiation of the double bubbles in Fig.~\ref{Fig:Bubbles} gives the
selfenergy loop contribution to the particle propagators as we show in 
Figs.~\ref{Fig:Pion-Self},\ref{Fig:Sigma-Self}. We can picture the dressed
propagator with a thick line which appears as a sum of the bare 
propagator (not shown in the figures) and the selfenergy 
contributions. Actually one can see these two figures as
a graphical representation of the gap equations 
for the dressed propagators. As we have 
mentioned already, at finite temperature and in the real time formalism, the propagators
appear to consist of two parts, a zero temperature part and a thermal part. We denote
each zero tempearture loop with continuous line and each thermal with a cut. Our choice
to keep only the non divergent parts of the integrals, is equivalent to taking
into account only the thermal propagators. This means that for pions, we are taking into account only 
the thermal self energy loops in Figs.~\ref{Fig:Pion-Self}b, d while 
for sigma those in Figs.~\ref{Fig:Sigma-Self}b,d.

Using a compact notation, the finite  
temperature effective potential can be written in the  form 
\begin{eqnarray} 
V(\phi,M)&=&\frac{1}{2}m^{2}\phi^{2}+\frac{1}{24}\lambda\phi^{4} 
\nonumber\\  
&& + Q_{\beta}(M_{\sigma}) + 3Q_{\beta}(M_{\pi})
\nonumber\\ 
&& - \frac{\lambda}{8}[F_{\beta}(M_{\sigma})]^{2} -\frac{5\lambda}{8}[F_{\beta}(M_{\pi})]^{2}
\nonumber\\  
&& - \frac{\lambda}{4}F_{\beta}(M_{\sigma})F_{\beta}(M_{\pi})~, 
\end{eqnarray} 
where in this last step, we have used the  
gap equations given by Eq.~(\ref{Eq:System-H}). The exact expressions for  
$F_{\beta}(M)$ and $Q_{\beta}(M)$, are 
given by Eq.~(\ref{Eq:FM}) and Eq.~(\ref{Eq:QM}) respectively. 

\subsubsection{High temperature limit}
  
In order to calculate the effective masses as functions of 
temperature, we need to solve the system of  the three equations given by Eq.~(\ref{Eq:System-H}) and  
Eq.~(\ref{Eq:Phi-H}). We first observe that if $\phi=0$, which happens  
in the high temperature 
phase, the two equations become degenerate, the particles have  
the same mass, and we have to solve only one equation  
\begin{equation} 
M^{2}=m^{2}+\lambda F_{\beta}(M)~. 
\label{Eq:Gap-H}
\end{equation} 
As in the expression of the effective potential, we 
keep only the finite temperature part of the integral. This last  
equation can be used to define a ``transition temperature''  
$T_{c1}$. This temperature is defined as the  
temperature, where both particles become massless.  Recall now  
that $F_{\beta}(M)$ is given by 
\begin{equation} 
\label{Eq:FBM} 
F_{\beta}(M)=\frac{T^{2}}{2\pi^{2}}\int_{0}^{\infty}\frac{x^2\, dx} 
{a^{1/2}} 
\frac{1}{\exp(a^{1/2})-1}~, 
\end{equation} 
where $a=x^{2}+y^{2}$ and  $y=M/T$. When the mass  
of the particles vanishes, this integral   
reduces to the well known result
\begin{equation} 
I(x)=\int_{0}^{\infty}\frac{x\, dx}{e^{x}-1}=\frac{\pi^{2}}{6}~. 
\label{Eq:pi/6} 
\end{equation}
Therefore by using Eqs.~(\ref{Eq:Gap-H}), (\ref{Eq:FBM}) and (\ref{Eq:pi/6}), we find that 
\begin{equation} 
0=m^{2}+\lambda\,\frac{T^{2}}{2\pi^{2}}\,\frac{\pi^{2}}{6}~. 
\end{equation}
Actually this defines the transition temperature $T=T_{c1}$ as 
\begin{equation} 
T_{c1}=\sqrt{2}\bigg(-\frac{ 6 m^{2}}{\lambda}\bigg )^{1/2}~. 
\end{equation} 
But, in  defining our model parameters, we have chosen that at zero 
temperature  
$\phi^{2}=f_{\pi}^{2}= -6 m^{2}/\lambda$, where  
$f_{\pi}=93\; \rm MeV$  is the  
pion decay constant, so we find that 
$T_{c1}=\sqrt{2} f_{\pi} \approx 131.5\; \rm MeV$.  

\subsubsection{Low temperature limit}

\begin{figure}
\includegraphics[scale=0.54]{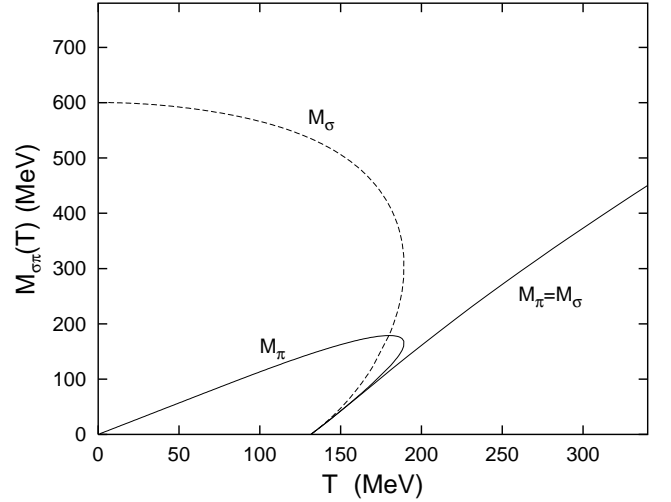} 
\caption{\label{hchirala}Solution of the system of gap equations in the chiral limit, $\varepsilon=0$. The 
sigma and pion effective masses are given as functions 
of temperature.}
\end{figure}

In the low temperature phase, we can 
eliminate  $\phi$ between Eqs.~(\ref{Eq:System-H}) and  
(\ref{Eq:Phi-H}). We then end up  
with the following nonlinear system  
\begin{eqnarray} 
M_{\sigma}^{2} &=& -2m^{2}-\lambda F_{\beta}(M_{\sigma}) 
-\lambda F_{\beta}(M_{\pi}) 
\nonumber\\ 
M_{\pi}^{2} &=& -\frac{\lambda}{3}F_{\beta}(M_{\sigma}) 
+\frac{\lambda}{3}F_{\beta}(M_{\pi})~. 
\label{Eq:System1} 
\end{eqnarray}
where, as before, we have ignored the divergent parts of $F(M_{\sigma/\pi})$. This system 
has been solved numerically using a Newton--Raphson method, and the solution 
is presented in Fig.~\ref{hchirala}~.
 
\noindent 
As shown in Fig.~\ref{hchirala}, the temperature $T_{c1}$ which  
is calculated numerically, was found to be  in excellent agreement with  
the value  obtained by using the limit of  
the  high temperature equations with degenerate masses.  
 
\begin{figure}
\includegraphics[scale=0.5]{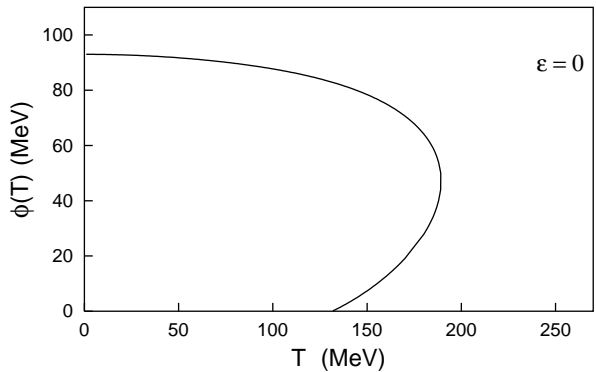} 
\caption{Evolution of the order parameter $\phi$ 
as a function of temperature.}
\label{hchiralb} 
\end{figure}

At this point, we can observe that  
there is an indication of a first order phase transition, because  
combining Eq.~(\ref{Eq:System-H.a}) with 
Eq.~(\ref{Eq:Phi-H}), we find that 
\begin{equation} 
M_{\sigma}^{2}=\frac{1}{3}\lambda\phi^{2}~. 
\end{equation} 
This last equation shows, of course, that the  
order parameter varies with temperature proportionally 
to the sigma mass. The temperature dependence of $\phi=\phi(T)$, is 
calculated by using the sigma mass as it was found 
by solving the system in Eq.~(\ref{Eq:System1}). This is shown in 
Fig.~\ref{hchiralb}, where it is obvious that this approximation predicts 
a first order phase transition, because $\phi$, which is the order parameter
of the phase transition, appears to have two different values for the same
temperature.

This last observation coincides with  
the qualitative picture given 
by Baym and Grinstein, in their early paper \cite{Baym:1977qb}, where their 
``modified Hartree approximation'' predicts a first order phase 
transition, as well. However, in contrast to our  
approximation, they had included quantum  
fluctuations  into their analysis as well. Signals of a first  
order phase transition have also 
been reported  in recent analyses by Randrup \cite{Randrup:1997hk}, Roh 
and Matsui \cite{Roh:1998ek}, and  Rischke and Lenaghan
\cite{Lenaghan:2000si}. However, this result seems to disagree
with other investigations of the linear sigma model (including 
fermions) or the Nambu--Jona--Lasinio model, where a second--order
transition has been reported \cite{Bailin:1985ak,Cleymans:kh,Bilic:1995ic}~.

\subsection{Evolution of the thermal potential} 

\begin{figure}
\includegraphics[scale=0.45]{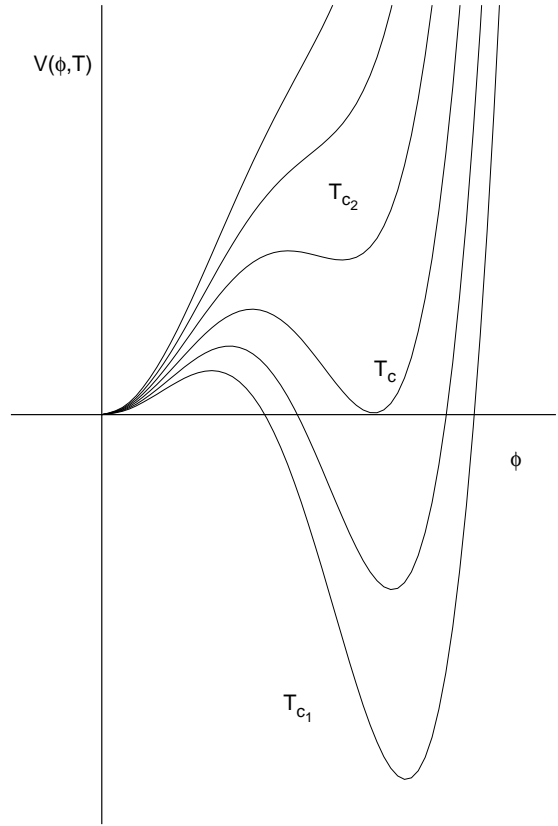} 

\caption{\label{Fig:Veff}A qualitative picture of the evolution of the 
effective potential $V(\phi,T)$ as a function of  
the order parameter $\phi$ for several temperatures. For our choice of parameters of the 
linear sigma model,  the two minima appear  
as degenerate at $T_{c}\approx 182\; {\rm MeV}$, while the upper spinodal point
(not shown explicitly) appears at  $T_{c_2}\approx 187\; {\rm MeV}$~.}
\end{figure}

In order to get more insight into the nature of the phase transition  
and verify that the transition is of the first order, we can 
calculate the effective potential $V(\phi,T)$ as a function 
of the temperature and the order parameter. So, we first solve  
numerically  the system of three equations in
Eq.~(\ref{Eq:System-H}) and Eq.~(\ref{Eq:Phi-H}) (where of course we keep only 
the finite temperature part of the integrals), and calculate the  
effective masses  
of the particles as functions of the order parameter and  
the temperature. Finally, the effective potential is calculated 
numerically, using these masses. The evolution of the potential  
for several temperatures 
is given in  Fig.~\ref{Fig:Veff}. The 
shape of the potential confirms that  
a first--order phase transition takes place, since it exhibits two  
degenerate minima at a  
temperature $T_{c}\approx 182\;{\rm MeV}$, which is usually defined 
as the transition  
temperature. The second minimum of  
the potential at $\phi \neq 0$ disappears at a temperature   
$T_{c2}\approx 187\;{\rm MeV}$~.

Normally, in first order phase transitions there are also two other temperatures of 
interest, apart from the transition temperature, but the 
relevant isotherms are not shown explicitly in this picture. These temperatures
$T_{c1}$  and $T_{c2}$ are called in condensed matter terminology, the lower 
and upper spinodal  
points, respectively. Between these temperatures, metastable  
states exist, and the system can exhibit supercooling or superheating. For 
$T_{c1}< T < T_{c}$, the metastable states are centred around the  
origin, since for $T_{c}< T < T_{c2}$ the metastable states occur 
for $\phi \neq 0$. When the system reaches $T_{c1}$ or 
$T_{c2}$, the curvature of the 
potential at the metastable minima vanishes. A discussion about  
first--order phase  
transitions, and more details about how these 
transitions proceed, can be found in  
\cite{Linde:1979px,Linde:1990}.

\subsection{The broken symmetry case $\varepsilon \neq 0$} 
 
\begin{figure}
\includegraphics[scale=0.54]{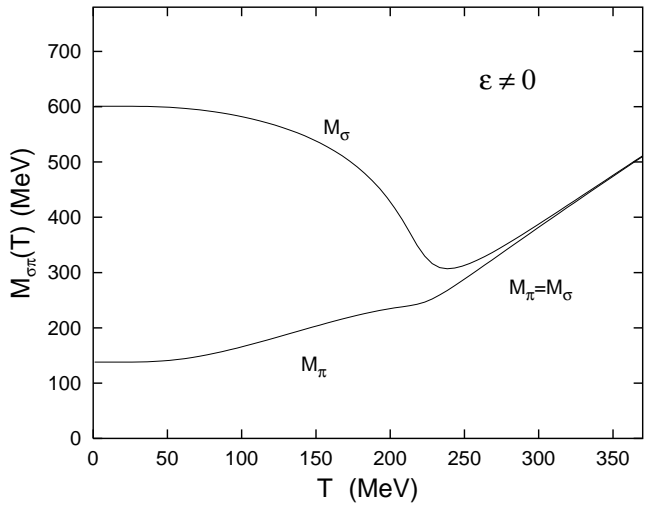} 

\caption{\label{Fig:hbreak}Solution of the system of gap equations in the case  
when $\varepsilon \neq 0$. At low temperatures the pions appear with 
the observed masses $m_\pi \approx 138 \,\textrm{MeV}$ .} 
\end{figure}

When $\varepsilon \neq 0$, the term linear in the sigma field  
into the Lagrangian  
generates the pion observed masses. This term  
is $G$ independent and so, minimization of the potential with 
respect to ``dressed'' propagators 
will give us the same set of gap equations for the effective 
masses as before. However, minimizing the potential with respect to $\phi$, we 
get the following equation 
\begin{equation} 
\left [ m^{2}+\frac{1}{6}\lambda\phi^{2}+\frac{\lambda}{2}F_{\beta}(M_{\sigma}) 
+\frac{\lambda}{2}F_{\beta}(M_{\pi})\right ]\phi-\varepsilon =0~. 
\label{Eq:Phi-H-B} 
\end{equation} 
In order to proceed, we need to solve the nonlinear system of  three 
equations in Eq.~(\ref{Eq:System-H}), and Eq.~(\ref{Eq:Phi-H-B}). We first observe that 
at $T=0$, Eq.~(\ref{Eq:Phi-H-B}) becomes 
\begin{equation} 
M_{\pi}^{2}=m^{2}+\frac{1}{6}\lambda\phi^{2}=\frac{\varepsilon}{\phi} 
=m_{\pi}^{2}~, 
\end{equation} 
where $m_{\pi}$ is the tree level pion mass. Then for 
$\phi=f_{\pi}$, we recover  
the relation between the pion mass at zero temperature and the  
symmetry breaking  
factor $\varepsilon$: $\varepsilon=f_{\pi}m_{\pi}^{2}$, where 
$f_{\pi}$ is the pion decay constant. 
We solved the system of Eq.~(\ref{Eq:System-H}) and Eq.~(\ref{Eq:Phi-H-B}) 
numerically, and the solution is presented  
in Fig.~\ref{Fig:hbreak}~.

At low temperatures, the pions appear with 
the observed masses, but their mass increases with temperature 
since the sigma mass decreases. At high temperatures (higher 
than $\sim 300\,\textrm{MeV}$), due to interactions in the thermal  
bath, all particles appear to have the  
same effective mass.

\begin{figure}
\includegraphics[scale=0.5]{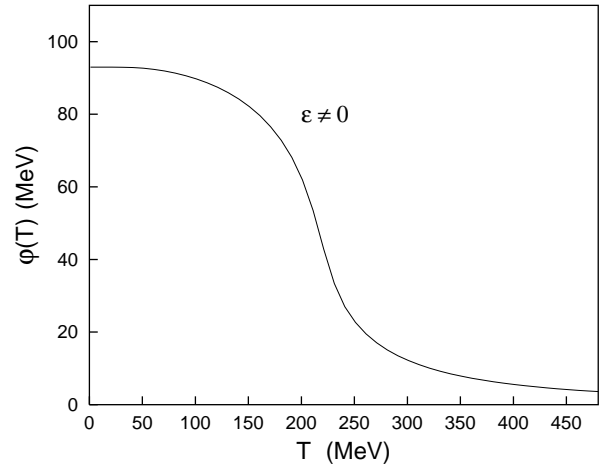} 

\caption{\label{Fig:hbreakb}Evolution of the order parameter $\phi$ as a function  
of temperature.} 
\end{figure}            
 
The presence of the symmetry--breaking term into  
the Lagrangian, modifies the evolution of the 
order parameter $\phi$, as well. As it is obvious in Fig.~\ref{Fig:hbreakb}, as the  
temperature increases, the order parameter decreases, and 
at very high temperatures vanishes smoothly. But in this case, the change is 
not a phase transition any more. We rather 
encounter a smooth crossover from a low--temperature 
phase, where the particles 
appear with different masses, to a high--temperature phase, where the 
thermal contribution to the effective masses makes them degenerate.

\section{\label{Sect:LargeN}Large $N$ approximation}

\subsection{Introduction}  
  
The large--$N$ approximation of the linear sigma model has been studied  
recently by Amelino--Camelia \cite{Amelino-Camelia:1997dd}, and our 
expressions are very  
similar to the ones obtained there, since the same method is used   
in both cases. However, in our approach, we do not consider   
the renormalization of the  
model because in our approximation we take into account only finite  
temperature effects. Our analysis, is in a sense, complementary to that  
in \cite{Amelino-Camelia:1997dd}, since we solve the system of gap 
equations and  
consider the effects  of the symmetry breaking term (the last term in the   
Lagrangian given by Eq.~(\ref{Eq:Lagrangian-N})), which is omitted in 
reference \cite{Amelino-Camelia:1997dd}.   
The renormalization of the model is investigated in  
a recent publication by Rischke and Lenaghan  
\cite{Lenaghan:2000si}.  Another  recent treatment of the $O(N)$ model 
appears in \cite{Nemoto:1999qf}, where they use 
the CJT method, as well, in order 
to calculate the thermal effective potential. The $O(N)$ version of the 
linear sigma model is very popular in condensed 
matter studies and it is a very well--studied model. Other studies of this
model in particle 
physics context are those in \cite{Bochkarev:1996gi,Bardeen:1983st}.
  
The generalised version  
of the meson sector  
of the linear sigma model is called the $O(N)$ 
or vector model, and is based on a set of $N$  
real scalar fields. The $O(N)$ model Lagrangian  
can be written as 
\begin{equation} 
\mathcal{L}=\frac{1}{2}(\partial_{\mu}\bm{\Phi})^{2}- 
\frac{1}{2}m^{2}\bm{\Phi}^{2} 
-\frac{1}{6N}\lambda \bm{\Phi}^{4}-\varepsilon\sigma~,
\label{Eq:Lagrangian-N}  
\end{equation} 
and in the absence of the last term, it remains invariant  
under  $O(N)$ symmetry  
transformations 
for any $N\times N$ orthogonal matrix. Our choice of taking $m^2$
negative, results in a $N$--dimensional ``mexican hat'' potential.

In order for our notation to be consistent with applications to pion  
phenomenology, we can identify $\Phi_{1}$ with the $\sigma$ 
field and the remaining $N-1$ components as the pion  
fields, that is $\bm{\Phi}=(\sigma, \pi_{1},\ldots,\pi_{N-1})$. The  
last term, $\varepsilon \sigma$ in the above expression  
has been introduced in order to generate masses 
for the pions. When $N=4$, this model is exactly the linear sigma 
model of the previous section.   

We proceed by examining the $O(N)$ within two approximation schemes, as we 
did in the Hartree approach, in the previous section.

\subsection{The chiral limit $\varepsilon=0$}    

In order to  study the $O(N)$ model, we proceed in absolutely analogous
steps as in the Hartree case in Section~\ref{Sect:Hartree}. So, by  shifting 
the sigma  
field as $\sigma \rightarrow \sigma+\phi$, the tree level   
propagators  are  
\begin{eqnarray}  
&&D^{-1}_{\sigma}(\phi;k)= k^2 +m^2  
+ \frac{2 \lambda}{N}\phi^2 ~, \nonumber \\  
&&D^{-1}_{\pi}(\phi;k)= k^2 +m^2 +\frac{2 \lambda}{3N}\phi^2  ~.  
\end{eqnarray}  
Then the  effective potential  
at finite temperature will appear as  
\begin{eqnarray}  
V(\phi,M) &=&\frac{1}{2}m^{2}\phi^{2}+\frac{1}{6N}\lambda\phi^{4}   
+\frac{1}{2}\int_{\beta}\ln G^{-1}_{\sigma}(\phi;k) \nonumber \\  
& & +\frac{N-1}{2}\int_{\beta}\ln G^{-1}_{\pi}(\phi;k)\nonumber \\   
& & +\frac{1}{2}\int_{\beta}  
[\mathcal{D}_{\sigma}^{-1}(\phi;k)G_{\sigma}(\phi;k)-1] \nonumber \\  
& & +\frac{N-1}{2}\int_{\beta}  
[\mathcal{D}_{\pi}^{-1}(\phi;k)G_{\pi}(\phi;k)-(N-1)]\nonumber \\  
& & +V_2(\phi,G_{\sigma},G_{\pi})~,  
\end{eqnarray}  
where the last term originates from the double bubble diagrams   
and its  contribution is  
\begin{eqnarray*}  
V_2(\phi,G_{\sigma},G_{\pi})&=&  
3\frac{\lambda}{6N}\left [  
\int_{\beta}G_{\sigma}(\phi;k)\right ]^{2} \nonumber \\   
& & +\frac{\lambda(N^{2}-1)}{6N}\left [  
\int_{\beta} G_{\pi}(\phi;k)\right ]^{2}\nonumber \\  
& & +\frac{\lambda(N-1)}{6N}\int_{\beta} G_{\sigma}(\phi;k)  
\int_{\beta}G_{\pi}(\phi;k)~.  
\end{eqnarray*}  
The weight factors appearing in the above expression  can be understood  
in a similar way as in the $O(4)$ case, with the only difference being  
the $N-1$ pion fields. Of course, it is easy to see that we recover the   
previous case by simply substituting $N=4$.  
  
As in the case of $\lambda\phi^{4}$ and the $O(4)$ model, we 
minimise  the effective  
potential with respect to the dressed propagators, and we get   
a set of gap   
equations. By using the same form    
for the dressed propagators as before, we end   
up with the following set of nonlinear gap  
equations for the thermal effective particle masses 
\begin{subequations} 
\begin{eqnarray*}
M_{\sigma}^{2}&=&m^2_{\sigma}
+\frac{2\lambda}{N}F_{\beta}(M_{\sigma})  
+\frac{2\lambda(N-1)}{3N}F_{\beta}(M_{\pi})
\\  
M_{\pi}^{2}&=&m^2_{\sigma}
+\frac{2\lambda}{3N}F_{\beta}(M_{\sigma})  
+\frac{2\lambda(N+1)}{3N}F_{\beta}(M_{\pi})~,
\end{eqnarray*}
\end{subequations}   
where we only keep the finite temperature part of   
the integrals, and we have defined the particle
masses at zero temperature as
\begin{eqnarray}
m^2_{\sigma}&=&m^{2}+\frac{2\lambda}{N}\phi^{2} \\ \nonumber
m^2_{\pi}&=&m^{2}+\frac{2\lambda}{3 N}\phi^{2}~. 
\end{eqnarray}  
As it is easy to observe, for $N=4$ we obtain   
identical expressions for the system of gap  equations, as in 
the case of the $O(4)$ model. 
  
In the large--$N$ approximation, which means that we ignore  
terms of $O(1/N)$, the system of   
the last two equations reduces to  
\begin{eqnarray}  
M_{\sigma}^{2} &=& m^{2}+\frac{2\lambda}{N}\phi^{2}  
+\frac{2\lambda}{3}F_{\beta}(M_{\pi})  
\nonumber\\  
M_{\pi}^{2}&=&m^{2}+\frac{2\lambda}{3 N}\phi^{2}  
+\frac{2\lambda}{3}F_{\beta}(M_{\pi})~.  
\end{eqnarray}  
We have retained the terms  quadratic in $\phi$   
since $\phi$ depends on $N$ as $\phi^{2}=-3 N m^{2}/2\lambda$, and so 
these terms are of $O(1)$. In order to solve this  
system, and be in ``some contact'' with phenomenology in the chiral  
limit, we can set  
$N=4$. Then, the pions are massless and the sigma   
has a mass $M^{2}_{\sigma}=-2m^{2}$ at zero temperature. Now our system   
is written as
\begin{subequations}
\label{Eq:System-LargeN}   
\begin{eqnarray}  
M_{\sigma}^{2}&=&m^{2}+\frac{1}{2}\lambda\phi^{2}  
+\frac{2\lambda}{3}F_{\beta}(M_{\pi}) \label{Eq:System-LargeN.a}
\\   
M_{\pi}^{2}&=&m^{2}+\frac{1}{6}\lambda\phi^{2}  
+\frac{2\lambda}{3}F_{\beta}(M_{\pi})~. \label{Eq:System-LargeN.b} 
\end{eqnarray}
\end{subequations}

The effective potential  
will appear in the form  
\begin{eqnarray}  
V(\phi,M)&=&\frac{1}{2}m^{2}\phi^{2}+\frac{1}{6N}\lambda\phi^{4}  
+ \frac{1}{2}\int_{\beta}\ln(k^{2}+M_{\sigma}^{2})\nonumber \\  
& &+ \frac{(N-1)}{2}\int_{\beta}\ln(k^{2}+M_{\pi}^{2})\nonumber \\  
& &-\frac{1}{2}(M_{\sigma}^{2}-m^{2}  
-\frac{2\lambda}{N}\phi^{2})F_{\beta}(M_{\sigma}) \nonumber \\  
& &-\frac{N-1}{2}(M_{\pi}^{2}-m^{2}-\frac{2\lambda}{3N}\phi^{2}) F_{\beta}(M_{\pi})\nonumber \\   
&&+\frac{\lambda}{2N}[F_{\beta}(M_{\sigma})]^{2}
+\frac{\lambda(N^{2}-1)}{6N}[F_{\beta}(M_{\pi})]^{2}\nonumber \\   
& &+\frac{\lambda(N-1)}{3N}F_{\beta}(M_{\sigma})F_{\beta}(M_{\pi})~.  
\end{eqnarray}  

In order to solve the system for the thermal effective masses 
in Eq.~(\ref{Eq:System-LargeN}), we proceed  as in the Hartree  
approximation. At very high temperatures, the potential has only one  
minimum, that at $\phi=0$, and in this case, the two equations  
become degenerate  
\begin{equation}  
M_{\sigma}^{2}=M_{\pi}^{2}= M^{2}=m^{2}+\frac{2\lambda}{3}F_{\beta}(M)~.  
\end{equation}  
This last equation actually defines the critical   
temperature. $F_{\beta}(M)$ is given by the same expression, as 
in the $O(4)$ case. The mass of the particles vanishes   
at the critical temperature, so we can use the result given in Eq.~(\ref{Eq:FBM})  
to find that the critical temperature is at  
\begin{equation}  
T_{c}=\sqrt{3}\left (-\frac{6 m^{2}}{\lambda} \right )^{1/2}  
=\sqrt{3}f_{\pi}\approx 161\,\textrm{MeV}~.   
\end{equation}

Before proceeding to examine the low--temperature phase, we should   
make an observation which actually  
exposes the significant difference between the Hartree approximation   
in the $N=4$ case, and the large--$N$ approximation. Minimizing    
the potential with respect   
to $\phi$ gives  
\begin{eqnarray*}  
\frac{dV(\phi,M)}{d\phi}&=&\phi\bigg [m^{2}  
+\frac{2\lambda}{3 N}\phi^{2} \\ \nonumber 
&&+\frac{2\lambda}{3N}F_{\beta}(M_{\sigma})  
+\frac{2\lambda(N-1)}{3N}F_{\beta}(M_{\pi})\bigg ]=0~,  
\end{eqnarray*}  
which, in the large--$N$ approximation becomes  
\begin{equation}  
\frac{dV(\phi,M)}{d\phi}=\phi\left[m^{2}+\frac{\lambda}{6}\phi^{2}  
+\frac{2\lambda}{3}F_{\beta}(M_{\pi})\right]=0~.  
\end{equation}  
Combining this last equation with 
Eq.~(\ref{Eq:System-LargeN.b})  above, we observe that  
\begin{equation}  
\frac{dV(\phi,M)}{d\phi}=\phi M^{2}_{\pi}=0~.  
\end{equation}  
Therefore, the large--$N$ approximation    
implies that the pions should be massless, in the 
low--temperature phase in accordance  
with the Goldstone theorem.   
  
\begin{figure}
\includegraphics[scale=0.5]{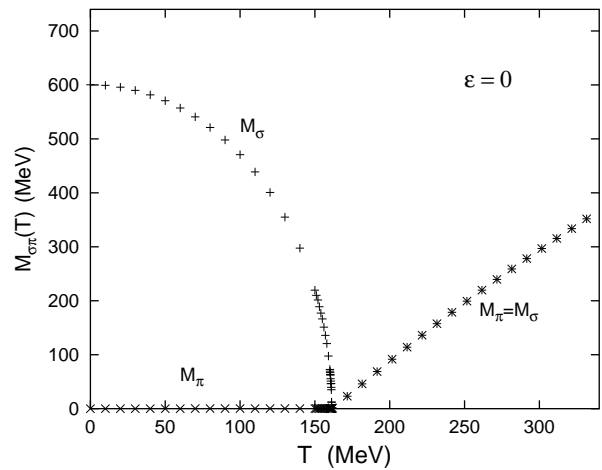} 

\caption{\label{Fig:nnchiral}Solution of the system of gap equations in the large--$N$  
approximation in the  chiral limit. At low temperatures, the pions 
appear as massless.}  
\end{figure}

\begin{figure}
\includegraphics[scale=0.5]{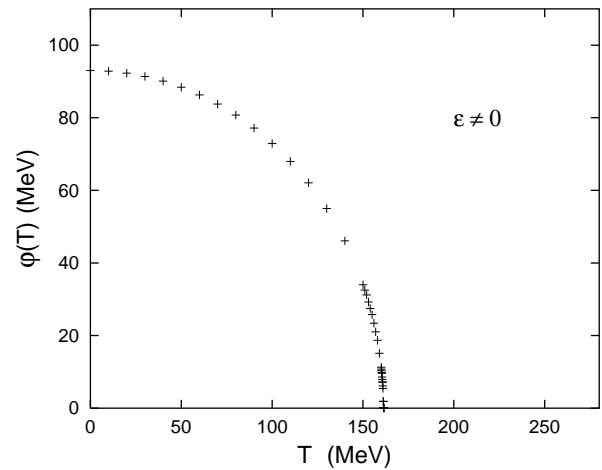} 

\caption{\label{Fig:nnchiralphi}Evolution of the order parameter with  
temperature in the large--$N$ approximation in the chiral limit $\varepsilon=0$}  
\end{figure}

This observation is reflected in the solution of the system   
of the gap equations, as is shown in Fig.~\ref{Fig:nnchiral}. The pions at   
low temperatures appear   
as massless, but at high temperatures the thermal contribution   
to the effective masses  
make them degenerate with the sigma. The order  
parameter vanishes continuously in this case as  
is shown in Fig.~\ref{Fig:nnchiralphi}, and  
this corresponds to a second--order phase transition.  

\subsection{The broken symmetry case $\varepsilon \neq 0$}  
  
As already mentioned  for the $O(4)$ case, the   
symmetry breaking   
term $\varepsilon\sigma$ has been   
introduced into the Lagrangian in order to generate the   
observed masses of the pions. The same can be done  
for the $O(N)$ model -- the only difference being the  
$N-1$ pion fields. Inserting this term into the expression   
for the effective potential and, differentiating with respect to $\phi$ we   
obtain, as in the $O(4)$ case, one more equation. In  
the large--$N$ approximation this is written as  
\begin{equation}  
\left [ m^{2}+\frac{1}{6}\lambda\phi^{2}  
+\frac{2\lambda}{3}(F_{\beta}(M_{\pi}) \right ]\phi-\varepsilon=0~.  
\label{Eq:LargeN-Phi}  
\end{equation}

\begin{figure}
\includegraphics[scale=0.50]{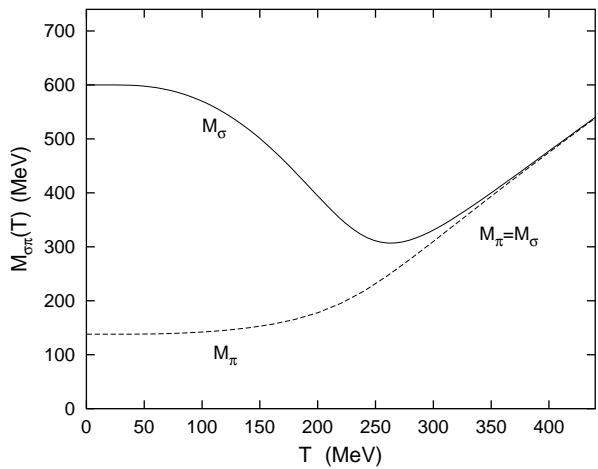}  

\caption{\label{Fig:nbreaka}Solution of the system of gap equations in the large--$N$  
approximation in the case of broken chiral symmetry  
$\varepsilon \neq 0$. At low temperatures, the pions appear with  
the observed masses.} 
\end{figure}

\begin{figure}
\includegraphics[scale=0.50]{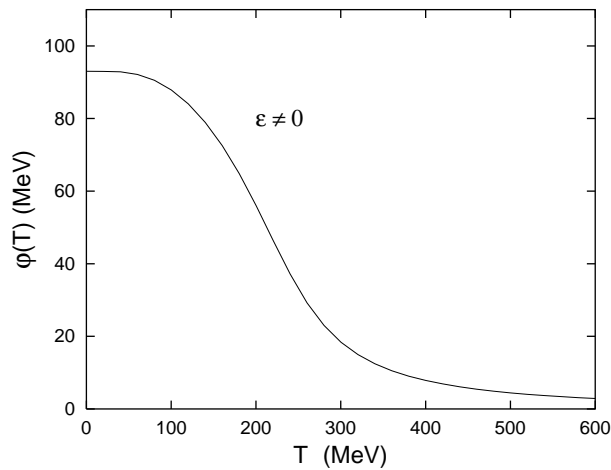}  

\caption{\label{Fig:nbreakb}Evolution of the order parameter $\phi$ as a function of   
temperature.}
\end{figure}

We have solved this last system of three equations given by  
Eq.~(\ref{Eq:System-LargeN}) and Eq.~(\ref{Eq:LargeN-Phi})  
numerically, and the solution is given in  
Fig.~\ref{Fig:nbreaka}~. As in the $N=4$ case, there is 
no longer any phase transition. We encounter again the   
crossover phenomenon between  
the low-- and high--temperature phases, the difference now being  
that the change of the order parameter  
(Fig.~\ref{Fig:nbreakb}) in the transition  
region is much smoother than  
the  ``sharper'' behaviour seen   
in the $N=4$ case, in Fig.~\ref{Fig:hbreakb}~.   
   
At this point, we would like to comment about the results 
presented in the previous two sections. Part of this work 
has also been presented elsewhere \cite{Petropoulos:1999gt,Petropoulos:1998en}. Firstly, the CJT 
formalism of composite 
operators proved to be very handy because we 
actually needed 
to calculate only one type of diagram. In both cases, we 
solved the system
of gap equations numerically, and found  the evolution with temperature of 
the effective thermal masses. In the Hartree 
approximation, we find a first--order phase 
transition  but, in contrast, the large--$N$
approximation predicts a second--order phase transition. This
last observation seems to be in agreement with
different approaches to the chiral phase transition, based 
on the argument that the linear sigma model belongs
in the same  universality class as other models which
are known  to exhibit second--order phase 
transitions \cite{Rajagopal:1993ah}. The same conclusion 
appears to be in the work of Bochkarev and 
Kapusta \cite{Bochkarev:1996gi}, where the linear
sigma model examined in the large--$N$ approximation, and they 
report second--order phase transition, as well. 

However, in the large--$N$ approximation, the sigma contribution is ignored and
this of course introduces errors when we
calculate  the critical temperature. In the case $N=4$, which 
is closer to phenomenology, we could have probably obtained 
a better approximation if we had
considered the effects of interactions given by the last two terms
in the Lagrangian given by Eq.~(\ref{Eq:Lint}). We attempt to study the 
effects of these
terms in Section~\ref{Sect:Beyond}. 

\begin{figure}
\includegraphics[scale=0.4]{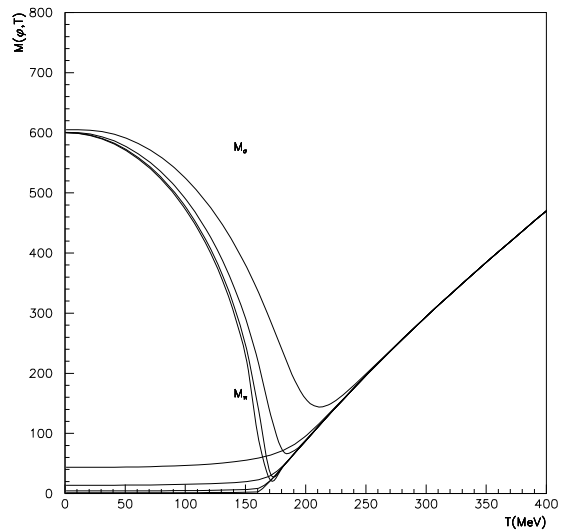}  

\caption{ Solution of the system of gap equations in the large--$N$  
approximation in the case of broken chiral symmetry  for various values of 
$\varepsilon $. We observe in this graph that as $\varepsilon \rightarrow 0$ 
the solution approaches the one given in Fig.~\ref{Fig:nnchiral}~. }
\label{eee}    
\end{figure}
\begin{figure}
\includegraphics[scale=0.5]{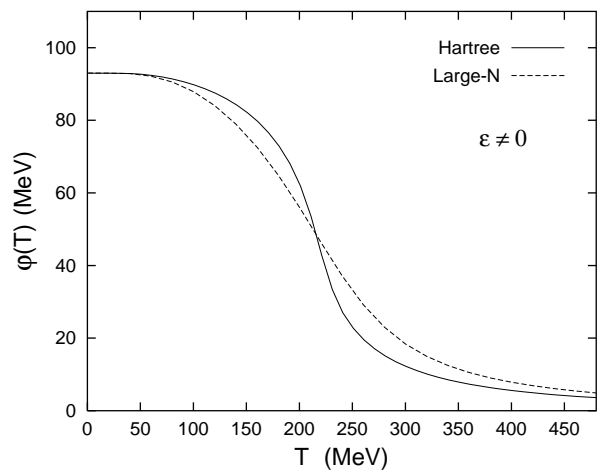}  

\caption{ The behaviour of the order parameter $\phi$ as a function of temperature 
in the case when chiral symmetry is broken ($\varepsilon \neq 0$) in Hartree (continuous line)
and large--N (dashed line).}
\label{phi-Hartree+N}    
\end{figure}

  
When we include the symmetry breaking term $\varepsilon \sigma$
which generates the pion observed masses, both in Hartree and 
large--$N$ approximations, we found that there is no longer any phase
transition. Instead, we observe a crossover phenomenon where the 
change of the order parameter in the Hartree case occurs more rapidly
in contrast to the smoother behaviour  exhibited in the large--$N$ 
approximation. The difference in the behaviour of the order parameter
in these two cases is shown in Fig.~\ref{phi-Hartree+N}~.   

 This observation confirms
results reported elsewhere as for example in the report by 
Smilga \cite{Smilga:1997cm}, or in the recent papers by Chiku and Hatsuda 
\cite{Chiku:1998va,Chiku:1998kd}. In the latter  analysis they also report
indication of a first--order phase transition in the chiral limit. In a 
recent investigation of the linear sigma model
by Rischke and Lenaghan \cite{Lenaghan:2000si}, the 
results  obtained are similar to ours, but 
they examine renormalization of the model as well.

\section{\label{Sect:Beyond}Beyond the Hartree approximation}  

\subsection{The sunset diagrams}
 
In order to understand the nature of the chiral
phase transition, in Section~\ref{Sect:Hartree}, we have  
calculated the thermal
effective potential in the Hartree approximation, which means that 
we did not take into account all the interactions stemming from the 
Lagrangian of the model. We found that this approximation predicts 
a first--order phase transition. However, this partial resummation 
of the Hartree approximation seems to conflict with other 
approaches, as for example, the one of Rajagopal 
and Wilczek \cite{Rajagopal:1995bc,Rajagopal:1993ah,Rajagopal:1993qz}, where 
the same model is investigated on the basis of universality arguments 
and a second--order phase transition is found. In order to reach 
deeper insight into the nature of the chiral phase transition, we 
need take into account all the interactions that the 
Lagrangian of the model describes, at least at the two--loop 
level calculation of the thermal effective potential. The study 
of the effective potential beyond leading order for the electroweak phase
transition has been studied by Arnold and Espinosa \cite{Arnold:1992rz}
while the QCD case is discussed in \cite{Arnold:ps,Arnold:1994eb}.
 
As it is shown in Section~\ref{Sect:First}, by shifting the sigma field 
as $\sigma\longrightarrow\sigma+\phi$, the 
interaction Lagrangian is given by  Eq.~(\ref{Eq:Lint}). This  
interaction part describes  proceses of the form 
$\pi \pi \rightarrow \pi \pi $, 
$\sigma \sigma \rightarrow \sigma \sigma$ and 
$\sigma \sigma \rightarrow \pi \pi$ or 
$\sigma \pi \rightarrow \sigma \pi$. We have considered the 
interactions of this form involving thermal 
pions and sigmas, when we calculated the bubble diagrams in the 
Section~\ref{Sect:Hartree}. However, the last two terms in Eq.~(\ref{Eq:Lint}) give rise to
processes of the form 
$\sigma \sigma \rightarrow \sigma \rightarrow \sigma \sigma$ and 
also to $\pi \pi \rightarrow \sigma \rightarrow \pi \pi $ as 
well as interactions between pions with a sigma exchange.

\begin{figure}
\includegraphics[scale=0.6]{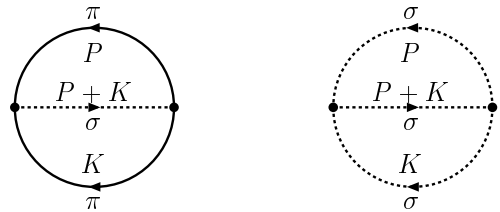}  
\caption{\label{Fig:Sunset} The two sunset--type diagrams which contribute to 
the effective potential of the linear sigma model at 
two loop level.} 
\end{figure}

In calculating the potential at the two--loop level, one now is dealing 
with the so--called ``sunset diagram''. There are two sunset diagrams, and
we show them in Fig.~\ref{Fig:Sunset}. The thermal effective potential 
now contains terms of the Hartree 
approximation (the double bubble diagrams) as in Section~\ref{Sect:Hartree}, plus the new terms 
of the sunset diagrams 
and it has the form 
\begin{equation} 
V_2(\phi,T)=V_{\textrm{Hartree}}+V_{\textrm{sunset}}~.  
\end{equation}
   
\subsection{The gap equations} 

\begin{figure}
\includegraphics[scale=0.63]{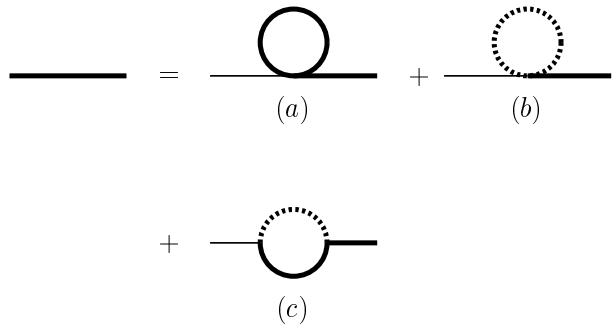}  
\caption{\label{Fig:pion-self-full}Schematic representation of the selfenergy contribution to pion
propagator. Continuous lines correspond to pion propagator, and dashed 
lines to sigma. The thick line represents the full propagator.}
\end{figure}

\begin{figure}
\includegraphics[scale=0.63]{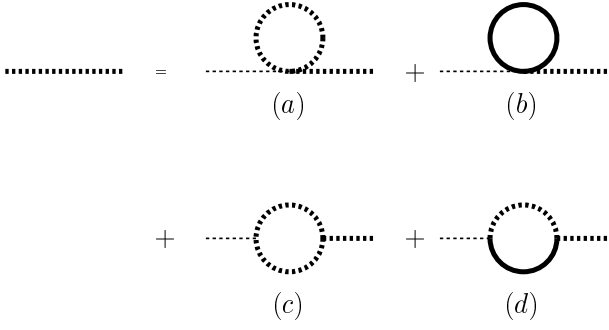}  
\caption{\label{Fig:sigma-self-full}Schematic representation of the self--energy contribution to sigma
propagator. Continuous lines correspond to pion propagator, and dashed 
lines to sigma. Thick line represents the full propagator.}
\end{figure}

As we have  already pointed out, functional 
minimization  
of the effective potential
with respect to the ``dressed propagator''  will give us a set of 
equations for the effective particle masses. Graphically, variation of 
the effective potential with respect to the dressed
propagator, corresponds to 
opening one propagator line in all diagrams of the effective 
potential. Taking the functional derivatives of the effective 
potential with respect the pion full propagator
$G_{\pi}$, we find the gap equation for the pions. We can represent 
this equation diagrammatically, as shown in 
Fig.~\ref{Fig:pion-self-full}~. Repeating this procedure for 
the sigma full propagator $G_{\sigma}$, we find 
the sigma gap equation. The diagrammatic 
representation for this gap equation is given in 
Fig.~\ref{Fig:sigma-self-full}~.

As discussed in the Section~\ref{Sect:First}, in the real time 
formalism, the thermal boson propagator 
becomes a two by two matrix, but at the level at which  we are working 
we need only the (1,1) component 
(we ommit the index 11 for simplicity) 
\cite{Leutwyler:1990uq,Banerjee:1991fu}. This 
is precisely the real time 
propagator, as given by  Dolan and Jackiw \cite{Dolan:1974qd}. This 
propagator consists of a sum of two parts: a zero 
temperature part which corresponds to

\begin{equation}
\label{Eq:real-prop-zero}  
\mathcal{D}_{0}(k)=\frac{i}{k^2-m^2+i\epsilon}
=i\textbf{P}\frac{1}{k^2-m^2}+\pi\delta(k^2-m^2)~, 
\end{equation} 

\noindent
where $\textbf{P}$ means the principal value, and a thermal part given by
\begin{equation}
\label{Eq:real-prop-thermal} 
\mathcal{D}_{T}(k,T)=2\pi \delta(k^2-m^2) n(k_0)~,  
\end{equation}
\noindent
where $n(k_0)$ is the Bose--Einstein distribution function 
\begin{equation} 
n(k_0)=\frac{1}{\exp(\beta\;|k_0|)-1}~, 
\end{equation} 
where the factror $\beta=1/T$ defines the inverse temperature while  
$k_0=\sqrt{\textbf{k}^2+m^2}$ is the energy. 

The first part of the propagator is the usual virtual particle 
exchange, present for all four--momenta $k$. The second term, the thermal 
part,  of the propagator describes 
real (on shell) particles existing in the hot 
plasma. These particle particles are present only when $k^2=m^2$ as the 
delta fuction constrains \cite{Rivers:1994rz}. In what follows, we denote 
the zero temperature part with continuous lines, and the thermal
part with lines incorporating a cut `` $|$ ''.

In this part of the work, we cannot follow blindly  the same recipe as in 
the Hartree case using a
dressed propagator. The reason is that now the self--energy 
is momentum dependent as it is obvious from the
graph  in Fig.~\ref{Fig:pion-self-full}c, and 
the ones in Fig.~\ref{Fig:sigma-self-full}c,d~. We can avoid this 
difficulty, if we adopt a ``Hartree like'' recipe and accept a form of 
the dressed
propagator as
\begin{equation} 
G_{\pi/\sigma}(\phi,M)=\frac{1}{K^2-M^2_{\pi/\sigma}}~,  
\end{equation} 
where we will consider this effective mass as obtained from the self--energy for 
zero external momenta.

\begin{figure}
\includegraphics[scale=0.6]{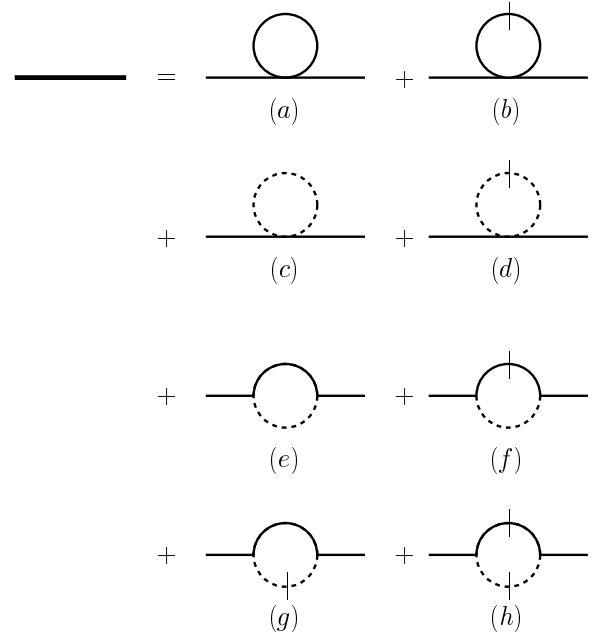}  
\caption{\label{Fig:pion-full} Schematic representation of the self--energy contribution  to pion
propagator. Continuous lines represent pions, dashed lines are 
sigmas, and lines
with vertical dashes correspond to the thermal propagators.}
\end{figure}

\begin{figure}
\includegraphics[scale=0.6]{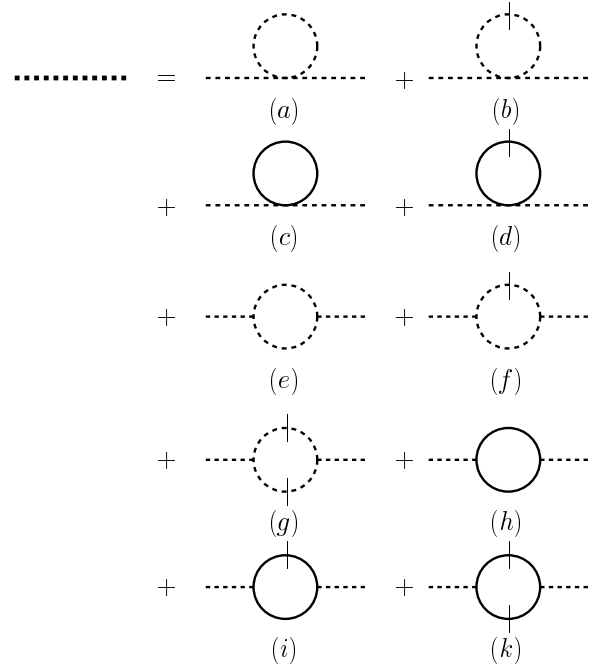}  
\caption{\label{Fig:sigma-full} Schematic representation of the self--energy contribution  to sigma
propagator. Continuous lines represent pions, dashed lines are sigmas, and 
lines with vertical dashes correspond to thermal propagators.}
\end{figure}

If we try to distinguish the thermal from  the 
quantum fluctuations, the pion full propagator could
be represented by the following set of graphs as is shown 
in Fig.~\ref{Fig:pion-full}~. On the other hand the full sigma propagator 
will be represented as is shown 
in the Fig.~\ref{Fig:sigma-full}~.

\subsection{The pion gap equation}

In our attempt to extend the Hartree approximation, our guide 
is the result of the low--energy theorem which was  presented 
in the first chapter. We have shown that the invariant amplitude 
of the sum of all 
four diagrams in Fig.~\ref{Fig:Tree-Graphs}  vanishes. This acts as 
motivation to consider the equivalent in the case where we deal with 
thermal particles. As a first step, we concetrate at the low--temperature 
region and mainly on the effects of the sunset diagrams 
on the effective masses of the pions. 

\begin{figure}
\includegraphics[scale=0.5]{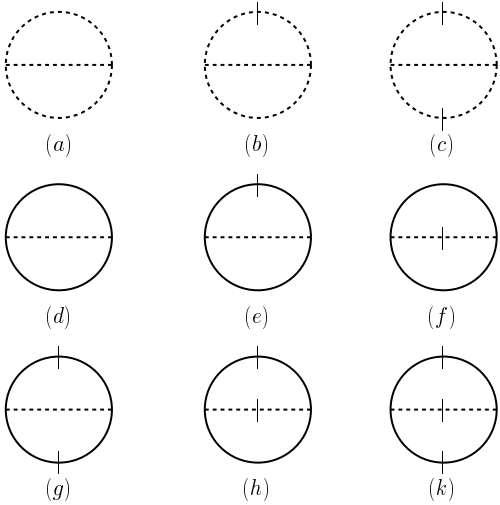}  
\caption{\label{Fig:Sunset-Full} Schematic representation of the sunset graphs. Continuous 
lines represent pions, dashed lines are sigmas, and lines
with vertical dashes correspond to thermal propagators.}
\end{figure}

We have shown the sunset diagrams which contribute to the effective potential
in Fig.~\ref{Fig:Sunset}. We can redraw these diagrams indicating the 
thermal propagators with a cut as we did in the graphic 
representation of the gap equations. These graphs should have the
same topology as in Fig.~\ref{Fig:Sunset}, but with one, two or three
cut lines representing thermal propagators. There is no graph 
with 3 thermal sigmas, because of energy conservation at the 
vertex. We show the complete set of sunset diagrams in 
Fig.~\ref{Fig:Sunset-Full}~. 

\begin{figure}
\includegraphics[scale=0.5]{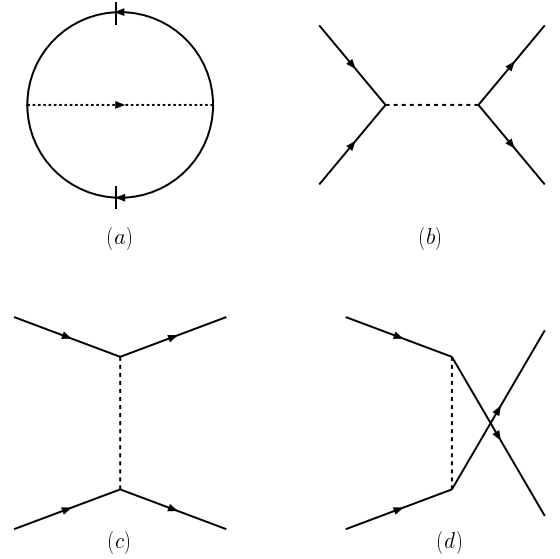}  
\caption{\label{Fig:equiv-graphs}(a) The sunset diagram with two thermal pions. A line with a 
dash corresponds to the thermal part of the full propagator. (b)-(d) The 
equivalent scattering diagrams involving thermal pions.}
\end{figure}

As a first approximation in the resummation of the subclass of the sunset 
diagrams, we only consider the sunset diagram with two thermal pions. Our 
motivation stems from the fact that at zero temperature and in the 
exact chiral limit, the pions 
being true Goldstone bosons have vanishing scattering 
amplitude as was presented in the first chapter. As is shown 
in  Fig.~\ref{Fig:equiv-graphs}, the sunset diagram 
with two thermal pions corresponds  to the
case where two real (thermal) pions collide to form a 
sigma, which afterwards decays into two pions. It also corresponds to
pion elastic scattering with sigma exchange.

The sunset diagram with two thermal pions is given in 
Fig.~\ref{Fig:equiv-graphs}a, and contributes the following term 
in the effective potential
\begin{equation}
V_{\textrm sunset}(\phi,G_{\sigma},G_{\pi})=\frac{\lambda^2\phi^2}{9}\mathcal{G}_{\pi}^{\theta}(\phi;k)
\mathcal{G}_{\pi}^{\theta}(\phi;p)\mathcal{G}_{\sigma}(\phi;k+p)
\end{equation}  
where $\mathcal{G}$ is an shorthand form of the integral
\begin{equation}
\mathcal{G}_{\pi/\sigma}(\phi;k)=\int_{\beta}G_{\pi/\sigma}(\phi;k)
\end{equation}  
Propagators indicated as $G^{\theta}$ correspond to
the thermal part of the real time propagator 
given by Eq.~(\ref{Eq:real-prop-thermal})~.

The full expession corresponding to the diagram in Fig.~\ref{Fig:equiv-graphs}a
is  
\begin{eqnarray}
I_{\sigma\pi_\theta\pi_\theta}(M_{\pi},M_{\sigma})&=&\int\frac{d^4 K}{(2 \pi)^4}\int\frac{d^4 P}{(2 \pi)^4}
2 \pi \,n(k_0)\;2 \pi \,n(p_0)  \nonumber \\
&&\times\frac{\delta(K^2-M_\pi^2)
\,\delta(P^2-M_\pi^2)}
{(P+K)^2-M_\sigma^2}~,
\end{eqnarray}
Following Arnold and Espinosa \cite{Arnold:1992rz}, we adopt the convenient notation $K^\mu=(k_0, \textbf{k})$,  for the  
four--momenta and, reserve the symbol $k=|\textbf{k}|$, for the magnitude  
of the three momentum. In this section we have adopted a different formalism than in 
Sections \ref{Sect:Hartree} and \ref{Sect:LargeN} so the superscripts denote to which particle's self--energy
we are reffered to, while the subscripts denote which particles are running in the loop. A subscript  
$\theta$ denotes a thermal propagator.

\begin{figure}
\includegraphics[scale=0.6]{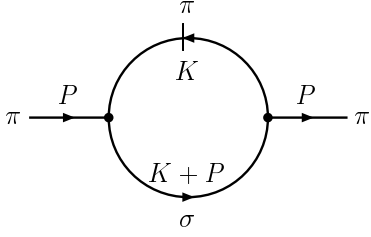}  
\caption{\label{Fig:pion-one-thermal} The pion self--energy diagram with one thermal pion.}
\end{figure}

Taking the derivatives of the potential with respect to the full 
propagator $G_{\pi}$, we get  
the  diagrammatic equation for the pion prapagator given in 
Fig.~\ref{Fig:pion-full}~. As a first approximation
we make  a selective summation by including the two graphs as 
in Hartree (Figs.~\ref{Fig:pion-full}b, d), plus 
the graph with one thermal pion (Fig.~\ref{Fig:pion-full}f). The topology 
of this last graph 
is given in Fig.~\ref{Fig:pion-one-thermal}~. 

If we use the real time form of the propagator as in 
Eqs.~(\ref{Eq:real-prop-zero}) and (\ref{Eq:real-prop-thermal}), the graph
in Fig.~\ref{Fig:pion-one-thermal} contributes with  
the following expression
\begin{equation} 
I^{\pi}_{\sigma \pi_\theta}(M_\pi,M_\sigma)=\int\frac{d^4 K}{(2 \pi)^4} 
\frac{2 \pi n(E_{\pi})\delta(K^2-M_{\pi}^2)} 
{(P+K)^2-M_\sigma^2+i\epsilon}~.
\end{equation} 
Then using Eq.~(\ref{Eq:real-prop-zero}),  we can split the above integral into real
and  imaginary parts. The imaginary part is related to dissipation 
phenomena that occur as the particles propagate within the thermal plasma.
We comment on these matters in Section~\ref{Sect:Others} where
we compare our approach to the chiral  
phase transition to the work of other investigators.

At the moment, we are interested only in the real part of the 
above integral, which has the form
\begin{equation} 
P(M_\pi,M_\sigma)=\int\frac{d^4 K}{(2 \pi)^4} 
\frac{2 \pi n(E_{\pi})\delta(K^2-M_{\pi}^2)} 
{(P+K)^2-M_\sigma^2}~.
\end{equation} 
Evaluation of this integral is given in Appendix~\ref{App:pion-se-graphs}. Then 
the resulting gap equation for the pions is 
\begin{eqnarray} 
M_{\pi}^{2}&=&m^{2}+\frac{1}{6}\lambda\phi^{2}+\frac{\lambda}{6}F(M_{\sigma}) 
+\frac{5\lambda}{6}F(M_{\pi})\nonumber\\ 
& &+\frac{\lambda^2\phi^2}{9}P(M_\pi,M_\sigma)~. 
\label{Eq:pion-new-gap} 
\end{eqnarray} 

Taking the derivative of potential with respect to the order parameter 
results in the 
following  equation 
\begin{eqnarray} 
0&=&m^{2}+\frac{1}{6}\lambda\phi^{2}+\frac{\lambda}{2}F(M_{\sigma}) 
+\frac{\lambda}{2}F(M_{\pi})\nonumber\\ 
& &+\frac{\lambda^2}{3}R(M_\pi,M_\sigma)~,
\end{eqnarray} 
where  the last term comes from the sunset diagram given 
in Fig.~\ref{Fig:equiv-graphs}a~.

Combining the above equations, we find that the thermal pion mass 
is given by
\begin{eqnarray} 
M_\pi^2 &=&-\frac{\lambda}{3}F(M_\sigma) 
+\frac{\lambda}{3}F(M_\pi)\nonumber\\ 
& &+\frac{\lambda^2 \phi^2}{9}P(M_\pi,M_\sigma) 
+\frac{\lambda^2}{3}R(M_\pi,M_\sigma)~. 
\end{eqnarray} 

As  a first step we are interested in  the low--temperature phase, and since sigma is very heavy, we do not
expect significant changes in the sigma mass, so we can use that $M^2_\sigma=\frac{1}{3}\lambda\phi^2$. Also the contribution 
from $F(M_\sigma)$  can be neglected as a first approximation, since it is 
exponentially supressed. Therefore in order to find out how the pion efective mass evolves with
temperature, we need to calculate the terms  $R$ and $P$. 

Evaluating 
the integral which corresponds to the 
pion self--energy graph in Fig.~\ref{Fig:pion-one-thermal}, we find
\begin{eqnarray} 
P(M_{\pi},M_{\sigma})&=& \frac{1}{2\pi^2}\int_0^\infty k^2\; dk
\frac{n(E_\pi^k)}{E_\pi^k} \nonumber\\
&&\times\frac{2M_\pi^2-M_\sigma^2}{(2 M_\pi^2-M_\sigma^2)^2
-4 M_\pi^2 (E_\sigma^k)^2}~.
\end{eqnarray}  
Details of this calculation can be found in Appendix \ref{App:pion-se-graphs}. On the 
other hand, evaluation of the sunset integral 
in Fig.~\ref{Fig:equiv-graphs}a, results in 
\begin{eqnarray} 
R(M_\pi,M_\sigma)&=&-\frac{1}{2(2\pi)^4}\int_0^\infty k\;dk\;g(k,M_\pi)\nonumber\\  
& &\times \int_0^\infty p\;dp\;g(p,M_\pi)\nonumber\\ 
          & &\times\ln\frac{X(k,p,M_\pi,M_\sigma)}{Y(k,p,M_\pi,M_\sigma)}
\end{eqnarray}
where
\begin{eqnarray*} 
X(k,p,M_\pi,M_\sigma)&=&M_\pi^2(k^2+p^2) + M^2_\sigma(M_\pi^2-M^2_\sigma/4)\nonumber\\ 
& & + (2 M^2_\pi-M^2_\sigma)k p~,
\end{eqnarray*}
 
\begin{eqnarray*}
Y(k,p,M_\pi,M_\sigma)&=&M_\pi^2(k^2+p^2) + M^2_\sigma(M_\pi^2-M^2_\sigma/4) \nonumber\\ 
& & - (2 M^2_\pi-M^2_\sigma)k p 
\end{eqnarray*} 
and we have used the abbreviatiated form for $g(p,M)$ introduced by Eq.~(\ref{Eq:g(z,a)}).  The exact 
calculation is given in Appendix~\ref{App:sunset-with-2p}~.

It is interesting to see how these integrals behave in the
limit of vanishing pion mass. Using a compact notation 
(introduced in Section~\ref{Sect:First}), $F(M)$ can be written as
\begin{equation}
F(M_{\pi/\sigma})=\frac{T^2}{2\pi^2}f(M_{\pi/\sigma}/T)~.
\end{equation}
\noindent
Rescaling the integral as $x=\beta k$, and in the limit of 
vanishing pion 
masses, we obtain the following expressions for the real part of the above 
integral, corresponding to self--energy graph 
\begin{eqnarray} 
P(M_{\pi},M_{\sigma})&=&-\frac{1}{M_\sigma^2}\frac{T^2}{2\pi^2} 
\int_0^\infty  g(x,M_\pi/T) \nonumber \\
&=&-\frac{1}{M_\sigma^2}\frac{T^2}{2\pi^2}f(M_\pi/T)~. 
\end{eqnarray} 
 
\noindent
On the other hand, the integral corresponding to sunset graph is given by 
\begin{eqnarray} 
R(M_{\pi},M_{\sigma})&=&-\frac{1}{2(2\pi)^2}\frac{8T^4}{M_\sigma^2}
\int_0^\infty x^2dx\; g(x,M_\pi/T)\nonumber \\
& &\times \int_0^\infty y^2dy\;g(y,M_\pi/T)~.
\end{eqnarray} 
\noindent
Recall now that the sigma mass is given by
\begin{equation}
M^2_\sigma=\frac{1}{3}\lambda\phi^2~. 
\end{equation} 
\noindent
Inserting these into the pion 
gap equation, we can observe that $F(M_{\pi})$ and
$P(M_{\pi},M_{\sigma})$ cancel exactly, so we end 
up with an expession where the thermal contribution to the 
pion mass squared is $\mathcal{O}(T^4)$. The same result under a different
approach has already been reported by Itoyama and Mueller
\cite{Itoyama:1983up}~.

\begin{figure}
\includegraphics[scale=0.45]{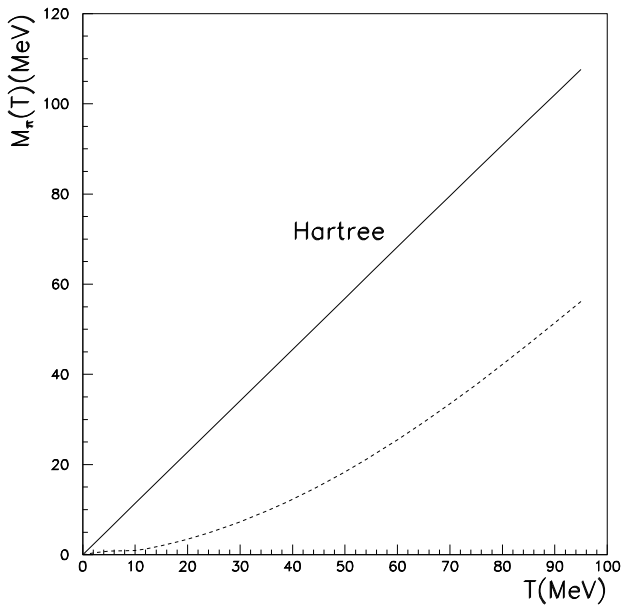}  
\caption{\label{Fig:apion} The low--temperature pion mass $M_{\pi}$, as a function of 
temperature. The continuous line corresponds to the Hartree approximation. }
\end{figure}

We illustrate the situation with in Fig.~\ref{Fig:apion}, where 
we plot the pion mass as we have calculated above, with the 
effective pion mass as was calculated in Hartree 
approximation. As we can observe 
in Fig.~\ref{Fig:apion}, in the Hartree calculation the 
pion mass is proportional to temperature. In contrast, inclusion 
of the self--energy graph results in the mass   being $\sim T^2$ and
obviously is small at low temperatures.  

\subsection{The sigma gap equation}

In our attempts to consider all the self--energy
graphs with one thermal particle 
in the loop, we faced serious numerical difficulties mainly 
from the pion self--energy graph. Being unable to solve
the complete system of equations, we have tried to 
understand the effects of each one graph individually.

\begin{figure}
\includegraphics[scale=0.6]{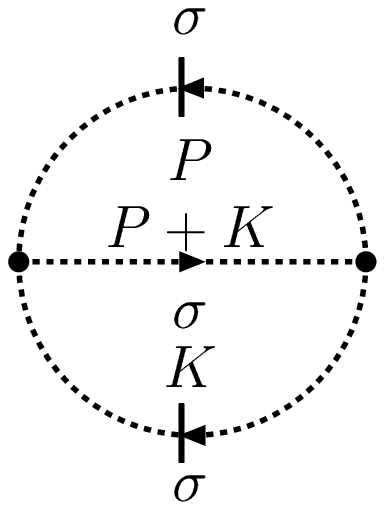}  
\caption{\label{Fig:sunset-2thermal-sigma}The sunset diagram with two thermal sigmas. A line with a
dash corresponds to the thermal part of the full propagator.}
\end{figure}

\begin{figure}
\includegraphics[scale=0.6]{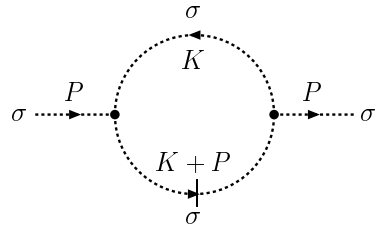}  
\caption{\label{Fig:sigma-self}The sigma self--energy diagram, with one thermal sigma.}
\end{figure}

If we add to the potential the sunset with two thermal sigmas, which 
is given  in the Fig.~\ref{Fig:sunset-2thermal-sigma}, differentiating, we
find a gap equation for sigma which contains the contribution of the sigma 
self--energy graph given in Fig.~\ref{Fig:sigma-self}.

\begin{figure}
\includegraphics[scale=0.45]{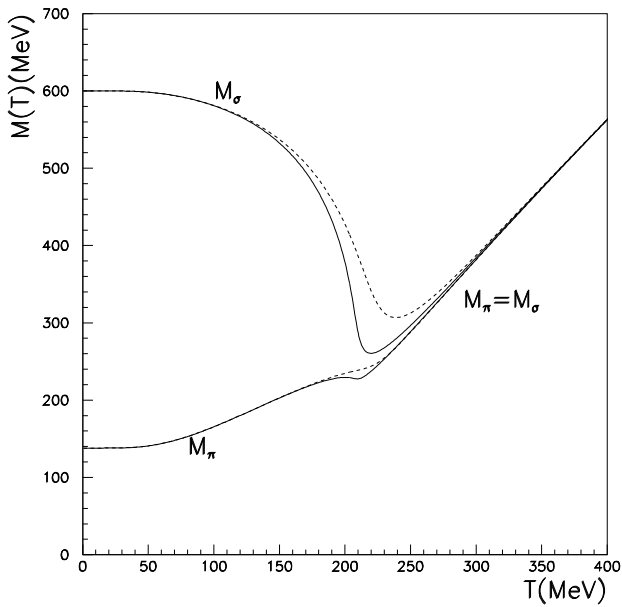}  
\caption{\label{Fig:sigma-pion-break}Solution of the system of gap equations for the sigma 
and pion effective masses. Dashed lines correspond to the Hartree solution
as in Section~\ref{Sect:Hartree}. The pion gap equation is as in the Hartree case, while
in the equation for the sigma we have included the sigma self--energy graph with one
thermal sigma.}  
\end{figure}

We proceed, using 
the same system of equations as in Hartree, with
the only difference being that the sigma gap
equation will contain the contribution of this graph. We have
also introduced the symmetry breaking term, so the pions 
are massive. The result, given in Fig.~\ref{Fig:sigma-pion-break}, is somehow the
one expected. We know that the sigma is heavy so, at low temperatures, there 
is no significant difference with the Hartree calculation. On
the other hand, at high temperatures the masses will approach ideal
gas behaviour. Only in the intermediate region of temperatures
we can observe  a small deviation of the Hartree result.

\begin{figure}
\includegraphics[scale=0.6]{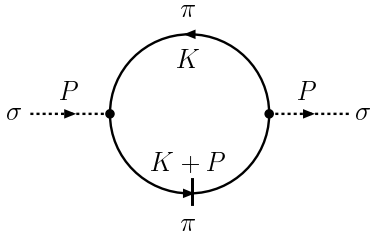}  
\caption{\label{Fig:sigma-self-2p}The sigma self--energy diagram with one thermal pion.}
\end{figure}

\begin{figure}
\includegraphics[scale=0.45]{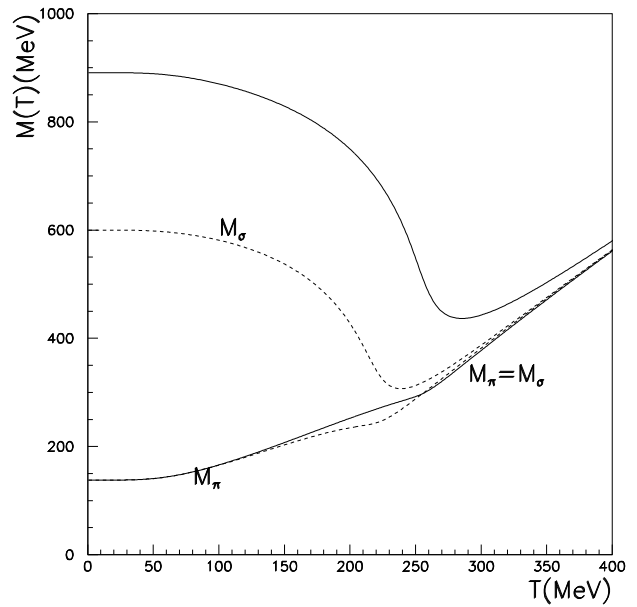}  
\caption{\label{Fig:sigma-pion-break2}Solution of the system of gap equations for the sigma 
and pion effective masses. Dashed lines correspond to Hartree solution
as in Section~\ref{Sect:Hartree}. The pion gap equation is as in the Hartree case, while
in  the equation for the  sigma we have included the sigma self--energy  graph with one
thermal pion.}  
\end{figure}

There is also no problem in  calculating the sigma self--energy
graph with one thermal pion given in Fig.~\ref{Fig:sigma-self-2p}. However, the 
addition of the relevant term into the sigma gap
equation deviates the sigma mass from $m_\sigma=600\;\textrm{MeV}$ as it 
is clear in Fig.~\ref{Fig:sigma-pion-break2}~. We
suspect that this is due to the fact that we do not include 
the vacuum graphs.

\section{\label{Sect:Others}Other approaches}

\subsection{Propagation of pions in hot plasmas}

Self--consistent approximations in many body systems have 
been tried a long time ago. To our knowledge, these
attempts date back to the 1960's, with early papers such as the
ones by Luttinger and Ward \cite{Luttinger:1960}, or the 
ones by Baym \cite{Baym:1962}. Recently 
these ideas have been employed by many people in various 
contexts as in the recent works by Knoll, Van Hees and others
\cite{vanHees:2000bp,Ivanov:2000ma}. 

Our approach to the chiral phase transition is not, of course, the only 
one. There have been a lot of studies on the subject since it is
related to  important phenomena as the formation of quark--gluon
plasma, the dilepton emmision etc. People have used various ways to approach
these problems, and we are going to outline their success here.  

A basic ingredient for the understanding of several physical processes 
that take place in hadronic plasmas, is the propagation 
properties of pions. The properties of pions in hot hadronic matter 
are encoded in the pion
propagator. The real part of the pion self energy 
is related to dispersion and group
velocity, while the imaginary part encodes the information about the pion 
absorption \cite{Goity:1989gs,Schenk:1993ru,Pisarski:1996mt}. In 
general, the expression which gives the pion mass shift is 
of the form 
\begin{equation}
(p^0)^2=\bm{p}^2+M_{\pi}^{2}+\Sigma(p^0,\bm{p})~,
\end{equation}
where $\Sigma$ is the pion self energy, and it depends on the physical
conditions of the medium, in which the pion propagates.
 
At the one loop level as it is the Hartree approximation 
there is no imaginary part in the self energy. The effective mass 
of the particles acquires  just thermal contribution. However, when we 
include the sunset diagrams into the potential
the mass gap equations contain terms which are momentum dependent. 

A recent work, which uses the linear sigma model in order to 
study dissipation phenomena in hot matter, is the one
by Ayala, Sahu and 
Napsuciale \cite{Ayala:2000ry,Ayala:2000px,Ayala:2000bx}~. Another work 
about dissipation properies of the pions, is the one by 
Rischke \cite{Rischke:1998qy}. The pion self--energy is calculated at the
one--loop level. In order to perform the study, the momentum running in
the loop is divided into soft and hard.  The model is used to study
disoriented chiral condensates.

\subsection{Other approaches using the sigma model}

The linear sigma model is a rich theory in its own right, however  it has 
been used 
as an effective theory to QCD transition since direct QCD calculations
are very complicated. The model has been used in various approaches 
of the QCD phase transition, both in and out of equilibrium. It has been 
especially popular in studies involving the formation of
disoriented chiral condensates. The formation of DCC was proposed
long ago, as a signal of the chiral phase transition \cite{Bjorken:1992xr}.

The CJT method and the linear sigma model have been used in other recent 
investigations, too. Our 
approach to the chiral phase transition appears to be complementary 
in most of these approaches. Throughout this report, we have not 
discussed the issue of renormalization. It is well known that
the linear sigma model is renormalizable at zero temperature. However, recent
attempts to incorporate renormalization at finite temperature, have not
always been successful. An attempt by 
Amelino--Camelia \cite{Amelino-Camelia:1997dd}, to examine renormalization 
of the linear
sigma model at  finite temperature results in violation of the
Goldstone theorem. We have mentioned this already in the third 
section. A recent successful treatment of the renormalization 
matters, is the one by Rischke and Lenaghan \cite{Lenaghan:2000si}.

In the Hartree approximation, we have calculated the effective 
masses of sigma and the pions. We call them effective 
masses, since they correspond to the modification 
of the bare particle masses, which result after their
``in medium '' interactions. Of course, they do not
correspond to the real observed masses, since in our 
approach, we have ignored quantum fluctuations 
throughout this work. 
  
In the work of Chiku and Hatsuda \cite{Chiku:1998va,Chiku:1998kd}, the 
optimal perturbation theory 
is employed in order to perform the calculations. They obtained
gap equations for the effective masses, which go further than   our Hartree
approximation. They include quantum fluctuations as well. So we
cannot really compare our results in full. However, they do not calculate the 
effective potential.

In the work of Nemoto, Naito and Oka \cite{Nemoto:1999qf}, there are many 
similarities to our approach. First of all, the CJT method is used for 
the calculations, however they renormalize the model, and
use a definition for the particle masses as the minimum
of the potential. 

In the work of Caldas, Mota and Nemes \cite{Caldas:2000ic}, the 
main difference is that they do
include fermions as well. Their calculation of evolution of
the condensate look very similar to our result. The show a first--order 
phase transition in the chiral limit,  and a crossover
when there is an explicit breaking of chiral symmetry. In the work of Bilic 
and Nicolic \cite{Bilic:1995ic}, they
also include fermions, and find a first--order phase transition.

Bochkarev and Kapusta \cite{Bochkarev:1996gi}, have focused in 
the differences between 
the linear and non--linear sigma model. They use the $O(N)$ version, and 
their result is similar to our large--$N$ approximation, since
they found a second--order phase transition as well. They also find
that the non--linear sigma model has a second--order phase 
transition as well.

\section{\label{Sect:Summary}Summary of the results}

\subsection{Conclusions}

We have studied 
the chiral phase transition
using the linear sigma model. In order to understand 
the nature of  the phase transition and how it could proceed, in the first 
two sections of this work, we have 
calculated the finite temperature effective
potential of this model in the Hartree and large--$N$ 
approximations using the CJT formalism of composite 
operators. This method has the advantage that we actually only need 
to calculate one type of diagram.

In both cases, we 
solved the system
of gap equations numerically, and found  the evolution with temperature of 
the thermal effective masses. In the Hartree 
approximation,  we find a first--order phase 
transition  but, in contrast, the large $N$
approximation predicts a second--order phase transition. This
last observation seems to be in agreement with
different approaches to the chiral phase transition based 
on the argument that the linear sigma model belongs
in the same  universality class as other models, which
are known  to exhibit second order phase 
transitions \cite{Rajagopal:1993ah}. 

However, in the large--$N$ approximation, the sigma contribution is ignored and
this, of course, introduces errors when we
calculate  the critical temperature. We found that the large--$N$
approximation predicts a higher transition temperature than the 
Hartree one. However,  we have concluded  that 
this is an artefact of the calculation, since both Hartree and 
large--$N$ approximations should have the same high temperature limit. We
should recall that, we can calculate the transition temperature by considering 
the high temperature limit from the mass gap equation, since 
it is defined as the temperature where the particles 
become massless. As 
we have pointed out in Section~\ref{Sect:Hartree}, in both cases 
and at high temperatures the behaviour of the
pion--sigma system approches that of the ideal gas.

When we include the symmetry breaking term $\varepsilon \sigma$
which generates the observed pion masses, we found that there is 
no longer any phase
transition, both in Hartree and 
large--$N$ approximations. Instead, we observe a crossover 
phenomenon, where the 
change of the order parameter in the Hartree case occurs more rapidly
in contrast to the smoother behaviour  exhibited in the large--$N$ 
approximation.  Our observation confirms
results reported recently by Chiku 
and Hatsuda \cite{Chiku:1998va,Chiku:1998kd}
using the ideas of optimal perturbation theory. In their analysis 
they also report
indication of a first order phase transition in the chiral limit. This
observation is closer to the real world, since the pions
are only approximately Goldstone bosons. 

Of course, as we have already pointed out, the linear 
sigma model is only an approximation
to the real problem which is QCD, but the study of the 
chiral phase transition in the framework of this model
could be a helpful guide to how one could tackle the 
original problem, and get some insight in the physics 
involved.  For the calculations both in Hartree and 
large--$N$ approximations, we have used the imaginary time formalism
which is adequate for studies at thermal equilibrium 
but if one is interested in studies of the dynamics of the phase 
transition, the real time formalism seems to be more 
convenient \cite{Smilga:1997cm}. We did not study the 
system far from equilibrium, however the real time formalism 
would allow us to extend the investigation, and consider 
the dynamics of the system. In real time formalism, we  have the advantage 
that the propagator splits into two
parts from the beginning, so making it easier to calculate 
thermal corrections to the effective masses.

However, our attempt to go futher than the Hartree approximation 
has run into serious difficulties. It is not absolutely clear
which class of diagrams we should consider. Our motivation
to include only the sunset--type diagrams with 
two thermal particle only,  comes out  from conciderations 
of the low--energy theorem, and our percistence into 
symmetry principles. This is why we have not tried to renormalise 
the model as well. It
is well known that the linear sigma model is a renormalizable 
theory at zero temperature, and that finite temperature 
effects do not indroduce new divergencies -- at least at the 
ultraviolet --, since the temperature acts as a natural cut--off.
However, renormalization of the model is investigated in 
other recent investigations as it was mentioned already.

The fact that near the transition temperature, the effect 
of higher loops may become important, it is well known for a long time, and suggests 
that further investigation is needed. We would suggest that an 
extension of our two loop 
calculation is tractable, provided one takes into account 
the efffects of all the loops contributing at the two--loop
level, as a first step. However, it is not clear
how the CJT method could be used to include the effects
of higher loops. Selfconsistent approximations have been 
used for a long time, but it is not clear how they work
beyond the one loop level.

\subsection{Some recent results}

We conclude this work reviewing some recent progress on the subject. Since the work which
presented in the previous sections was initially written, there has has been quite 
some progress on similar studies. The linear sigma model 
has always been a very popular model in studies
attempting to mimic the most of the low energy region of QCD. 

Perhaps the most interesting  result, which is related to this work and has been published 
recently, is the work of Baacke and Michalski \cite{Baacke:2002pi}, where 
the authors examine the $O(N)$ linear sigma model beyond the Hartree  
approximation. The same authors have also studied the sigma model at non--equilibrium in \cite{Baacke:2001zt}
(see also \cite{Michalski:2003fp}). However, it seems that their approach does 
not suffer from the difficulties that
we have faced trying to solve the full system of gap equations at the two loop level of 
the effective potential. An  essential difference between our  approach and the one by  Baacke and Michalski is 
the renormalisation of the model. As we have seen already in Section~\ref{Sect:First} the 
CJT method is based on a resummation of the  2PI graphs. The analysis of \cite{Baacke:2002pi} is 
based on the a variation of the CJT method, which is called 2PPI resummation. This method has 
been initiated by Verschelde and Coppens \cite{Verschelde:bs} and it is presented in 
a number of papers \cite{Verschelde:2000dz,Verschelde:2000ta,Smet:2001un}.

In order to study the effective potential Baacke and Michalski have calculated  the full subset 
of sunset diagrams. This of course makes their approach more complete. However, our attempt was 
focused mainly on the sunset with two thermal pions trying to test if 
the low energy theorem has an analogy at finite temperature, rather than 
tackling issues like for example renormalization of the model at the full 
two loop calculation, or attempts to consider the full two loop case.

An  alternative approach to the problem of the selfconsistent approximations, is the work 
of Knoll and van Hees \cite{vanHees:2001ik,VanHees:2001pf,vanHees:2002bv} where they use
the $\Phi$ derivable approximation.  This method orinates from the work of Baym \cite{Baym:1962} and it has essentially 
many similarities with CJT. However, it seems that renormalisation is 
controlled better under this calculational scheme.

The CJT method is an attractive framework in order one to perform selfconsistent approximations and 
has attracted quite some attention. The method is used in a recent paper by Roder \textit{et al} \cite{Roder:2003uz}
where the authors study restoration of chiral symmetry and the low lying meson spectrum at finite 
temperature. Also in another  recent study  of the $O(4)$ linear sigma model Phat \textit{et al} \cite{Phat:2003eh} have used 
the CJT  method as well. In their analysis, in contrast to our approach, they have studied the issue of
renormalisation while they have incorporated the effects od higher loops. Other recent work on the $O(N)$ linear
sigma model, but without using CJT, is the 
work of Patkos \textit{et al} \cite{Patkos:2002ec,Patkos:2002vr,Patkos:2002vq,Patkos:2002xb} where they study 
the behaviour of the spectal functions, as well as, the position of the sigma pole on the 
complex energy plane. 

\begin{acknowledgments}
These are not the original acknowledgments
of the thesis, where I had to thank all the people who supported me during that period, and I do thank them
again but the list is too long to be repeated here. However, 
I would like, once again, to express my gratitude to Mike Birse who was my adviser for this work
for his constant support and encouraging. He was always there to offer his help and advise in 
any case. Joannis Papavassiliou now at Valencia University
has offered me lots of advise and help during his stay in Manchester. Ray Rivers who during the 
final examination of the thesis with his critical and well aimed  questions has really helped me to 
look deeper and to understand more about the physics which 
I was trying to understand. Finally I would like to thank Eef van Beveren
who offered me a job on his project and has encouraged me to use some time to update this work  and 
Alex Blin for the many discussions we had during my stay in Portugal. I would like also to thank 
EPSRC for supporting my PhD studies. Financial support of the  
{\it Funda\c{c}\~{a}o para a Ci\^{e}ncia e a Tecnologia}
of the {\it Minist\'{e}rio da
Ci\^{e}ncia e da Tecnologia} of Portugal, under  contract CERN/FIS/43697/2001 and
POCTI/FNU/49555/2002 during my present 
stay in Portugal is also acknowledged.
\end{acknowledgments}

\appendix

\section{\label{App:Potential}The potential at one--loop}

In this appendix, we review the calculation of  
the one--loop contribution to the effective potential for a $\lambda\phi^4$
theory, using the imaginary time formalism. This is a well presented
calculation, and appears in many research papers and textbooks, as for example 
references \cite{Dolan:1974qd,Kapusta:1989,Das:1997gg}. However, in 
order to make this presentation more self--contained, and also 
compare it with the real time calculation, we will reproduce the 
basic steps here. The propagator is given by 
\begin{equation} 
 \mathcal{D}^{-1}(\phi;k) = k^2 +m^2 ~. 
\end{equation}
The one loop contribution to the effective potential is of the form
\begin{eqnarray}
V(\phi,T) & = & 
 \int_\beta d^4 k \,\ln \mathcal{D}^{-1} ( k; \phi)  \nonumber \\ 
& = & - \frac{1}{2\beta}\sum_n \int \frac{ d^3\textbf{k} } {(2\pi)^3} 
\ln (k^2 + m^2) \nonumber \\ 
& = &  - \frac{1}{2\beta}\sum_n \int \frac{ d^3\textbf{k} } {(2\pi)^3} 
\ln \left(\frac{4 n^2\pi^2}{\beta^2}-E^2\right)~,
\end{eqnarray}
where  $E^2 = \textbf{k}^2 + m^2 $, and we have indicated the sum over 
the Matsubara frequency 
$k_0=\omega_n = 2\pi n \beta^{-1}$ explicitly. We should recall that 
$\beta=1/T$, where $T$ is the temperature.

\noindent
In order to evaluate this sum we can define
\begin{equation} 
u(E)=\sum_n \ln\left( \frac{4 n^2\pi^2}{\beta^2}+E^2\right) 
\end{equation} 
so the derivative is 
\begin{equation} 
\frac{\partial u(E)}{\partial E}=
\sum_n  \frac{2E}{4 n^2\pi^2/\beta^2+E^2}~.
\end{equation}
\noindent
We can now use the formula \cite{Gradshteyn:1980}
 \begin{equation}
\sum_n \frac{x}{x^2 + n^2}= -\frac{1}{2x}+\frac{1}{2}\pi\coth \pi x~,
\end{equation}
\noindent
to find that
\begin{equation} 
\frac{\partial u(E)}{\partial E}=
2\beta\left( \frac{1}{2}+\frac{1}{e^{\beta E}-1}\right)~, 
\end{equation}
and
\begin{equation} 
u(E)=2\beta\left( \frac{E}{2}+\frac{1}{\beta}\ln(1-e^{\beta E})\right)~.  
\end{equation}
The result is summarised as
\begin{eqnarray}
V(\phi,T) &=&  V^0 (\phi) + V^{\beta} (\phi) \\ \nonumber 
              & = &   \int \frac{ d^3\textbf{k} }{(2\pi)^3}\frac{E}{2}
+\frac{1}{\beta} \int \frac{ d^3\textbf{k} }{(2\pi)^3}
\ln ( 1 - e^{ - \beta  E} )~.
\end{eqnarray}

\section{One loop mass correction}

\begin{figure}[h]
\includegraphics[scale=0.45]{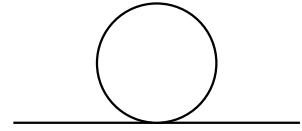}  
 
\caption{The self--energy contribution  to scalar
propagator.}
\label{1111}
\end{figure}

We have seen in n Section \ref{Sect:First}, that the one--loop correction 
to scalar propagator only causes a mass shift. The relevant 
graph is shown in Fig.~\ref{1111}~.

\noindent
This graph involves evaluation of the following integral
\begin{eqnarray}
F(m)& = & - \frac{1}{2\beta}\sum_n \int \frac{ d^3\textbf{k} } {(2\pi)^3}
\frac{1}{(2 n \pi/ \beta)^2+E^2}\\ \nonumber 
& = &  - \frac{1}{2\beta}\left(\frac{\beta}{2\pi}\right)^2
\sum_n \int \frac{ d^3\textbf{k} } {(2\pi)^3} 
\frac{1}{n^2+(\beta E/2\pi)^2}~.
\end{eqnarray}
\noindent
If we use the formula 
\begin{equation}
\sum_{n} \frac{1}{x^2 + n^2}= 
\frac{\pi}{x}+\coth \pi x~,
\end{equation}
we find that
\begin{equation}
F(m)= -\frac{1}{2}\int \frac{ d^3\textbf{k} } {(2\pi)^3} 
\frac{1}{2 E}\coth\left(\frac{\beta E}{2}\right)~. 
\end{equation}
From the fact that
\begin{equation}
\coth \beta x=1+2n(2x)~,
\end{equation}
where $n(E)$ is the Bose--Einstein  distribution function
\begin{equation}
n (E) = \frac{1}{ e^{ \beta E} - 1 }~, 
\end{equation}
we find that he above integral is written as a sum of two parts
\begin{eqnarray}
F(m)& = & F_0(m)+F_\beta(m)\\ \nonumber 
& = & \frac{1}{2}\int \frac{ d^3\textbf{k} } {(2\pi)^3} 
\frac{1}{2 E}+ \frac{1}{2}\int \frac{ d^3\textbf{k} } {(2\pi)^3} 
\frac{1}{E}\frac{1}{ e^{ \beta E} - 1 }~. 
\end{eqnarray}

\section{\label{App:Keldysh}The Keldysh contour}

The imaginary time or Matsubara formalism has the disadvantage that 
if one wants to calculate non equilibrium quantities, an analytic continuation 
to real time must be performed. In 
this formalism, we are integrating in the complex plane
of the Minkowski time, from $t=0$ to $t-i\beta$. The 
periodicity of the fields
enables us to generalise this interval to, $t_0$ to $t_0-i\beta$, for 
any real $t_0$. 

\begin{figure}[h]
\includegraphics[scale=0.57]{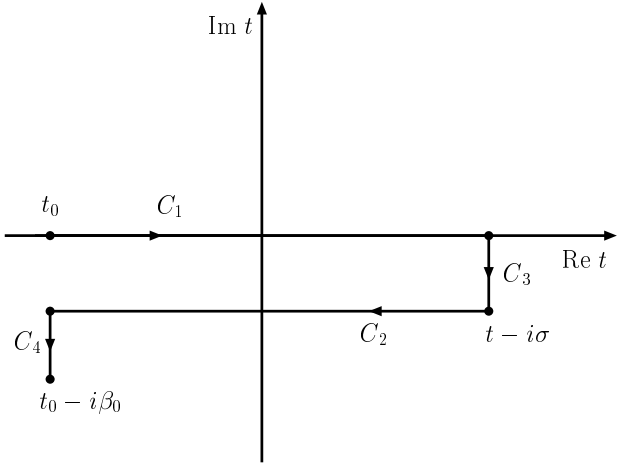}  

\caption{Keldysh contour}  
\label{Fig:Keldysh-Contour}
\end{figure}

Alternatively, one can work directly in real time. In this case, we need 
a different choice of contour 
with the same endpoints which will contain the
real time axis. A choice of  contour for
time integration, which is called the Keldysh contour,
is shown in Fig.~\ref{Fig:Keldysh-Contour}. 

As we can
see in Fig.~\ref{Fig:Keldysh-Contour}, the Keldysh contour consists  of four
pieces. However, there is
an alternative choice of contour \cite{Evans:1995gb}.  For analyticity reasons, it 
can be shown that the contributions 
from $C_3$ and $C_4$ can be neglected. Therefore, we are left with 
two possibilities for both the initial and final points of integration
to lie either on $C_1$ or $C_2$. This gives four different
propagators, which are usually written in matrix form. We 
have given these propagators in Section~\ref{Sect:Formalisms}.

\section{\label{App:pion-se-graphs}The pion self--energy graphs}

\subsection{\label{App:pion-se-graphs-tp}The pion self--energy graph with one thermal pion}

\begin{figure}
\includegraphics[scale=0.6]{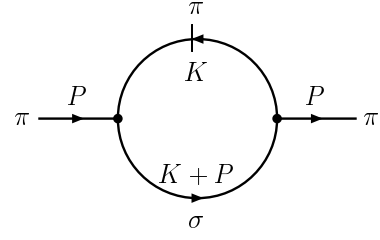}

\caption{\label{Fig:pion-self+pion-apped}The pion self--energy diagram with one thermal pion.}
\end{figure}

Although the calculation of the self--energy graphs 
for $\lambda\phi^4$  is a well known subject, and appears in many 
research papers, we repeat the derivation 
here, hoping to make the presentation
of the calculations more complete. Evaluation of the self--energy 
can be done using any of the formalisms:
imaginary time, real time or  thermo--field dynamics 
\cite{LeBellac:1996,Contreras:1990gi,Arnold:1993qy,Weldon:1983jn,Fujimoto:1986me}. We find 
the real time convenient at this stage. The pion self--energy graph 
with one thermal pion in the loop, is 
given in Fig.~\ref{Fig:pion-self+pion-apped}. 

As  mentioned already in Section~\ref{Sect:Beyond}, in our case, we need only the 
(1,1) component of the matrix propagator. This  has a zero temperature
part
\begin{equation}
\label{real-prop-zero-apped}  
\mathcal{D}_{0}(k)=\frac{i}{k^2-m^2+i\epsilon}
=i\textbf{P}\frac{1}{k^2-m^2}+\pi\delta(k^2-m^2)~, 
\end{equation} 

\noindent
where $\textbf{P}$ means the principal value, and a thermal part given by
\begin{equation}
\label{real-prop-thermal-apped} 
\mathcal{D}_{T}(k,T)=2\pi \delta(k^2-m^2) n(k_0)  
\end{equation}
\noindent
where $n(k_0)$ is the Bose--Einstein distribution function 
\begin{equation} 
n(k_0)=\frac{1}{\exp(\beta\;|k_0|)-1}~,
\end{equation} 
with $\beta=1/T$ the inverse temperature  
and $k_0=\sqrt{\textbf{k}^2+m^2}$. 

We use the following convention for 4--momenta 
$K^{\mu}=(k_0,\textbf{k})$, and reserve $k$ for the magnitude of 
the 3-momentum, so $k=|\textbf{k}|$. Since we are interested 
in the real part of the self--energy graph, we ignore the second 
part of the vacuum propagator (the one with 
the delta function). Then the exact 
expression corresponding to the graph in Fig.~\ref{Fig:pion-self+pion-apped} is written as
\begin{equation} 
I^{\pi}_{\sigma \pi_\theta}(M_\pi,M_\sigma)=\int\frac{d^4 K}{(2 \pi)^4}
\frac{2 \pi n(k_0)\delta(K^2-M_{\pi}^2)}
{(P+K)^2-M_\sigma^2}~.
\end{equation} 
As we have mentioned earlier, the superscripts denote to which particle's self--energy
we are reffered to, while the subscripts denote which particles are running in the loop. A subscript  
$\theta$ denotes a thermal propagator.

\noindent 
By using partial fractioning, we can write the above integrand as
\begin{eqnarray*}
\frac{1}{(P+K)^2-M_{\sigma}^2} &=&\frac{1}{2E_\sigma^{k+p} }
\bigg [ \frac{1}{p_0+k_0-E_\sigma^{k+p}} \nonumber \\
&&-\frac{1}{p_0+k_0+E_\sigma^{k+p}} \bigg ]~.
\end{eqnarray*} 
\noindent
Where the superscript in the  expression for the energy has the obvious meaning, $E^q_a=\sqrt{q^2+m^2_a}$.
In order to perform the integration, we use the following property of the Dirac delta function
\begin{equation}
\delta(x^2-a^2)=\frac{1}{2|a|}\bigg [ \delta(x+a)+\delta(x-a)\bigg ]~.
\label{Eq:delta}
\end{equation}
Inserting everything in the integral, and integrating over the 
delta function, results in

\begin{eqnarray*}
I^{\pi}_{\sigma \pi_\theta}(M_\pi,M_\sigma)&=&\frac{1}{2\pi^2}\int_0^\infty k^2 dk
\frac{n(E_\pi^k)}{2 E_\pi^k E_\sigma^{k+p}} \nonumber \\
&&\;\;\;\; \times \bigg [\frac{1}{p_0+E_\pi^k-E_\sigma^{k+p}}\nonumber \\
&&\;\;\;\;\; -\frac{1}{p_0+E_\pi^k+E_\sigma^{k+p}} \nonumber \\
&&\;\;\;\;\; +\frac{1}{p_0-E_\pi^p-E_\sigma^{k+p}} \nonumber \\
&& \,\;\;\;\;-\frac{1}{p_0-E_\pi^k+E_\sigma^{k+p} } \bigg ]~,
\end{eqnarray*}

\noindent
or, in a more compact form, as

\begin{eqnarray*}
I^{\pi}_{\sigma \pi_\theta}(M_\pi,M_\sigma)&=&\frac{1}{2\pi^2}\int_0^\infty k^2  dk
\frac{n(E_\pi^k)}{E_\pi^k}\nonumber \\
&&\times\bigg [ \frac{1}{(p_0+E_\pi^k)^2-(E_\sigma^{k+p})^2}\nonumber \\
&&\;\;\;\;+\frac{1}{(p_0-E_\pi^k)^2-(E_\sigma^{k+p})^2}\bigg ]~.
\end{eqnarray*}

In the case when we consider pions as static, then ${\textbf{p} = 0 }$ 
and so $p_0=M_\pi$, therefore we can write the above as

\begin{eqnarray}
I^{\pi}_{\sigma \pi_\theta}(M_\pi,M_\sigma)&=&\frac{1}{4\pi^2}\int_0^\infty k^2\; dk
\frac{n(E_\pi^k)}{E_\pi^k} \nonumber \\
&&\times\bigg [ \frac{1}{(M_\pi+E_\pi^k)^2-(E_\sigma^k)^2}\nonumber \\
&&\;\;\;\;+\frac{1}{(M_\pi-E_\pi^k)^2-(E_\sigma^k)^2}\bigg ]~,
\end{eqnarray}  
or even as 
\begin{eqnarray}
I^{\pi}_{\sigma \pi_\theta}(M_\pi,M_\sigma)&=&\frac{1}{2\pi^2}\int_0^\infty k^2\; dk
\frac{n(E_\pi^k)}{E_\pi^k} \nonumber \\
&&\times\frac{2M_\pi^2-M_\sigma^2}{(2 M_\pi^2-M_\sigma^2)^2
-4 M_\pi^2 (E_\sigma^k)^2}~.
\end{eqnarray}

\subsection{\label{App:pion-se-graphs-ts}The pion self--energy graph with one thermal sigma}  

\begin{figure}[h]
\includegraphics[scale=0.6]{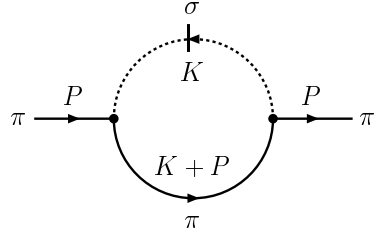}  

\caption{\label{Fig:pion-self+sigma-apped}The pion self-energy diagram with one thermal sigma.}
\end{figure}

Another graph which contributes to the pion 
self--energy, is the one given in Fig.~\ref{Fig:pion-self+sigma-apped}.   
As we can easily observe, apart from that the thermal particle 
is a sigma now, this graph is absolutely analogous to the one
presented above. Therefore, following the same procedure as before, we 
end up with the following expression

\begin{eqnarray}
I^{\pi}_{\pi\sigma_\theta}(M_{\pi},M_{\sigma})&=&\frac{1}{2\pi^2}\int_0^\infty k^2\; dk
\frac{n(E_\sigma^k)}{E_\sigma^k} \nonumber \\
&&\times\bigg [ \frac{1}{(M_\pi+E_\sigma^k)^2-(E_\pi^k)^2}\nonumber \\
&&~~+\frac{1}{(M_\pi-E_\sigma^k)^2-(E_\pi^k)^2}\bigg ]
\end{eqnarray}

\section{\label{App:sigma-se-graphs}The sigma self--energy graphs}

\subsection{\label{App:sigma-se-graphs-ts}The sigma self--energy graph with one thermal sigma}

Evaluation of the sigma self--energy graphs can be done in a completely analogous 
way as to one of the pions. However, since it involves identical particles it is 
even easier. Therefore, 

\begin{figure}[h]
\includegraphics[scale=0.6]{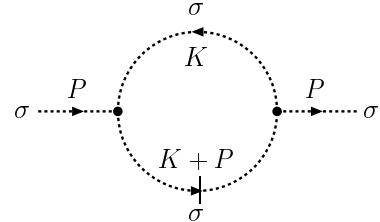}  
\caption{The sigma self--energy diagram with one thermal sigma.}
\end{figure}

\begin{equation}
I^{\sigma}_{\sigma \sigma_\theta}(M_{\sigma})=\frac{1}{2\pi^2}\int_0^\infty k^2\; dk
\frac{n(E_\sigma^k)}{E_\sigma}
\frac{1}{M_\sigma^2-4E_\sigma^2}~.
\end{equation}  

\subsection{\label{App:sigma-se-graphs-tp}The sigma self--energy graph with one thermal pion}
\begin{figure}[h]
\includegraphics[scale=0.6]{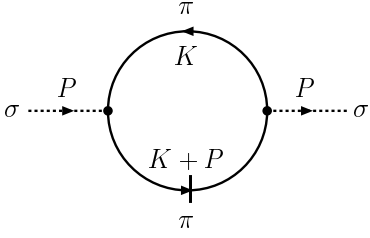}  
\caption{\label{Fig:sigma-self-1tp-app}The sigma self--energy diagram with one thermal pion.}
\end{figure}

In this case we have the topology as in Fig.~\ref{Fig:sigma-self-1tp-app}
Evaluating the relevant integral as above, we find
\begin{equation}
I^{\sigma}_{\sigma \pi_\theta}(M_{\pi})=\frac{1}{2\pi^2}\int_0^\infty k^2\; dk
\frac{n(E_\pi^k)}{E_\pi}
\frac{1}{M_\pi^2-4E_\pi^2}~.
\end{equation}

\section{\label{App:sunset-graphs}Calculation of sunset graphs}

\subsection{\label{App:sunset-with-2p}The sunset with two thermal pions}

The sunset diagrams consist of a set of diagrams with the topology 
given in Fig.~\ref{Fig:Sunset-Full}~. However, in the approximation adopted for this work, we  ignore 
quantum fluctuations, so we do not take into account the graphs 
with one thermal propagator or none. The graphs with three thermal propagators do not 
contribute (because of conservation of energy), so we are left 
only with the graphs containing two thermal propagators.

\begin{figure}[h]
\includegraphics[scale=0.6]{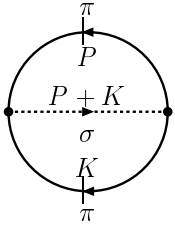} 
\caption{\label{Fig:sunset-2p}The sunset diagram, with two thermal pions. A line with a
dash corresponds to the thermal part of the full propagator.}
\end{figure} 

In the case of two thermal pions in the sunset as shown in Fig.~\ref{Fig:sunset-2p}, we deal with the expession
\begin{eqnarray}
I_{\sigma\pi_\theta\pi_\theta}(M_{\pi},M_{\sigma})&=&\int\frac{d^4 K}{(2 \pi)^4}\int\frac{d^4 P}{(2 \pi)^4}
2 \pi \,n(k_0)\;2 \pi \,n(p_0)  \nonumber \\
&&\times\frac{\delta(K^2-M_\pi^2)
\,\delta(P^2-M_\pi^2)}
{(P+K)^2-M_\sigma^2}~,
\label{Eq:Ispp}
\end{eqnarray}
and in order to evaluate this integral, we use the property of the delta function
given by Eq.~(\ref{Eq:delta}). Then our expression will consist of  four terms containing products
of delta functions of the form
\begin{equation}
\delta(k_0\pm E_\pi)\delta(p_0\pm E_\pi)~.
\label{Eq:combination}
\end{equation}
 
In order to perform this 
integration,  it is convenient to choose $\textbf{k}$  as
defining  the polar axis,  so then $\theta$ is the angle between 
the three vectors  $\textbf{k}$ and $\textbf{p}$. Then, the first term of 
the integral given in Eq.~(\ref{Eq:Ispp}) will appear as     

\begin{eqnarray*}
R_a(M_{\pi},M_{\sigma})&=&\frac{1}{2(2 \pi)^4}\int\frac{dk_0}{k_0}
\frac{k dk }{e^{\beta k_0 }-1} \nonumber \\
&&\times\int\frac{dp_0}{p_0}
\frac{p dp }{e^{\beta p_0 }-1}\;\mathcal{A}(k,p,\theta)
\end{eqnarray*}
where we have defined $\mathcal{A}=\mathcal{A}(k,p,\theta)$ as a shorthand of the lengthy
expression
\begin{equation*}
\mathcal{A}=\frac{k p \sin\theta\; d\theta\;\delta(k_0- E_\pi^k)\;\delta(p_0- E_\pi^p)}
{k_0^2-k^2+p_0^2-p^2+2 k_0p_0-2 kp\sin\theta-M_\sigma^2}~.
\end{equation*}
\noindent
After performing the delta function integration, we end up with the 
result
\begin{eqnarray*}
R_a(M_{\pi},M_{\sigma})&=&\frac{1}{2(2 \pi)^4}\int\frac{k dk }{E_{\pi}^k}
\frac{1 }{e^{\beta E_\pi^k }-1}\nonumber \\
&&\int\frac{p dp}{E_\pi^p}
\frac{1}{e^{\beta E_\pi^p }-1}k p \;\mathcal{B}(k,p,\theta)
\end{eqnarray*}
and we have defined again a shorthand expression $\mathcal{B}=\mathcal{B}(k,p,\theta)$ as 

\begin{equation*}
\mathcal{B}=\frac{k p \sin\theta\; d\theta}
{(E_\pi^k + E_\pi^p)^2-k^2-p^2-2 k p \sin\theta-M_\sigma^2}~.
\end{equation*}
\noindent

\noindent
Finally, performing the angular intergration, we find
\begin{eqnarray*}
R_a(M_{\pi},M_{\sigma})&=&\frac{1}{2(2 \pi)^4}\int\frac{k dk }{E_{\pi}^k}
n( E_\pi^k )\int\frac{p dp}{E_\pi^p}
n( E_\pi^p )\nonumber \\
&&\times\ln\frac{2\;M_\pi^2-M^2_\sigma+2 k_0\;p_0+2kp}
{2\;M_\pi^2-M^2_\sigma+2 k_0\;p_0-2kp}~.
\end{eqnarray*}

\noindent
In repeating this procedure, there will be four other similar terms, one
for each of the combinations of delta functions 
given by Eq.~(\ref{Eq:combination}). Then the final result can be written as

\begin{eqnarray}
I_{\sigma\pi_\theta\pi_\theta}(M_\pi,M_\sigma)&=&-\frac{1}{2(2 \pi)^4}\int\frac{k dk }{E_{\pi}^k}
n( E_\pi^k )\int\frac{p dp}{E_\pi^p}
n( E_\pi^p )\nonumber \\
          & &\times\ln\frac{A(k,p,M_\pi,M_\sigma)}{B(k,p,M_\pi,M_\sigma)}
\end{eqnarray}
where
\begin{eqnarray*} 
A(k,p,M_\pi,M_\sigma)&=&M_\pi^2(k^2+p^2) + M^2_\sigma(M_\pi^2-M^2_\sigma/4)\nonumber\\ 
& & + (2 M^2_\pi-M^2_\sigma)k p~,
\end{eqnarray*}
\begin{eqnarray*}
B(k,p,M_\pi,M_\sigma)&=&M_\pi^2(k^2+p^2) + M^2_\sigma(M_\pi^2-M^2_\sigma/4) \nonumber\\ 
& & - (2 M^2_\pi-M^2_\sigma)k p~. 
\end{eqnarray*} 

\subsection{\label{App:sunset-with-2s}The sunset with two thermal sigmas}

\begin{figure}[h]
\includegraphics[scale=0.6]{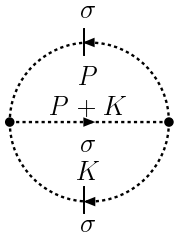} 

\caption{\label{Fig:sunset-2s}The sunset diagram, with two thermal sigmas. A line with a
dash corresponds to the thermal part of the full propagator.}
\end{figure}

In the case when we need to calculate the sunset with two 
thermal sigmas, the situation will be absolutely analogous to
the one presented above for the two pions. The diagram has the topology 
given in Fig.~\ref{Fig:sunset-2s}~.

Then the exact expression corresponding to this graph is given by

\begin{eqnarray}
I_{\sigma\sigma_\theta\sigma_\theta}(M_{\pi},M_{\sigma})
&=&\int\frac{d^4 K}{(2 \pi)^4}\int\frac{d^4 P}{(2 \pi)^4}\,
2 \pi n(k_0)\;2 \pi n(p_0) \nonumber \\
&&\times\frac{\delta(K^2-M_\sigma^2)\,\delta(P^2-M_\sigma^2)}
{(P+K)^2-M_\sigma^2}~.
\end{eqnarray}

In order to evaluate this, we repeat the steps as in case with the two thermal
pions. The final result will be of the form

\begin{eqnarray}
I_{\sigma\sigma_\theta\sigma_\theta}(M_{\pi},M_{\sigma})
&=&\frac{1}{2(2 \pi)^4}\int\frac{k dk }{E_{\sigma}^k}
n( E_\sigma^k )\int\frac{p dp}{E_\sigma^p}
n(E_\sigma^p ) \nonumber \\
          & &\times\ln\frac{4(k+p)^2+3M^2_\sigma+4kp}
            {4(k+p)^2+3M^2_\sigma-4kp}~.
\end{eqnarray}

\appendix
\end{document}